\documentclass[pra,showpacs,twocolumn]{revtex4}
\usepackage{amsmath,amssymb,epic,eepic}
\usepackage[dvips]{graphicx}
\usepackage{psfig,epsfig}
\usepackage{color}
\usepackage{isolatin1}

\begin{document}

\title{Entanglement and decoherence of $N$ atoms and a mesoscopic field in a cavity}

\author{T. Meunier}
\affiliation{Kavli Institute of Nanoscience, Delft University of Technology, P.O. 
Box 5046, 2600 GA Delft, The Netherlands} 
\author{A. Le Diffon}
\author{C. Ruef}
\author{P. Degiovanni}
\affiliation{CNRS-Laboratoire de Physique de l'Ecole Normale Sup{\'e}rieure de Lyon, %\\
46, All{\'e}e d'Italie, 69007 Lyon, France}
\author{J.-M. Raimond}
\affiliation{Laboratoire Kastler Brossel, Département de Physique de l'Ecole Normale Supérieure, 
24 rue Lhomond, F-75231 Paris Cedex 05, France}

%\today

\begin{abstract}
We investigate the behavior of $N$  atoms resonantly coupled to a single electromagnetic field 
mode sustained by a high quality cavity, containing a mesoscopic coherent field. We show 
with a simple effective hamiltonian model that the strong coupling between the cavity and 
the atoms produces an atom-field entangled state, involving  $N+1$ nearly-coherent components 
slowly rotating at different paces in the phase plane. The periodic overlap of these components 
results in a complex collapse and revival pattern for the Rabi oscillation. We study the influence 
of decoherence due to the finite cavity quality factor. We propose a simple analytical model, based 
on the Monte Carlo approach to relaxation. We compare its predictions with exact calculations 
and show that these interesting effects could realistically be observed on a two or three atoms sample 
in a 15 photons field with circular Rydberg atoms and superconducting cavities.
\end{abstract}


\maketitle

 \section{Introduction}
 
Cavity quantum electrodynamics experiments with circular Rydberg atoms and superconducting cavities 
are well suited for the realization of tests of fundamental quantum processes and of simple quantum 
information processing functions \cite{Raimond:2001-1}. They make it possible, in particular, 
to prepare mesoscopic quantum superpositions, made of coherent field components with different 
classical attributes (phase and amplitude). 
They have opened the way to studies of the decoherence dynamics on these states, 
at the quantum/classical boundary \cite{Brune:1996-2}. These early experiments, involving fields 
containing a few photons only, were based on the dispersive atom-field interaction. The atom, off 
resonance with the cavity mode, behaved as a state dependent 
transparent dielectrics modifying transiently the 
cavity frequency and, hence, the field phase. An atom in a superposition of levels produces then 
a quantum superposition of phase shifts, a situation reminiscent of the famous Schr\"odinger cat 
situation.

Much faster phase shifts can be realized through the resonant atom-cavity interaction. The complex 
Rabi oscillation phenomenon in a mesoscopic field results in an atom-field entanglement induced by 
photon graininess. The initially coherent cavity field is rapidly cast in a superposition of two components 
with different phases. This phase splitting is a mesoscopic effect that disappears in 
the classical limit of a very large field, which is then left unaffected by the atoms.
This resonant phase splitting effect has been evidenced for fields containing 
up to a few tens of photons \cite{Auffeves:2003-1}. Its coherence has been checked using an echo 
technique borrowed from NMR \cite{Meunier:2005-1}, following a proposal by 
Morigi {\it et al} \cite{Morigi:2002-1}. The resonant atom-field interaction thus 
opens the way to decoherence studies with large photon numbers.
These experiments focused on a simple situation with a single atom coupled to the cavity mode. 
Recent experimental advances \cite{Maioli:2005-1}
allow us to envision experiments with samples containing a well known number $N>1$ of atoms. 
They would merge the concepts of cavity QED with the atomic ensemble manipulations recently 
put forth for quantum information processing.
In this context, it is particularly interesting 
to study the resonant interaction of such a multi-atom sample 
with a mesoscopic field.

 \medskip
 
In this paper, we study the resonant interaction of an atomic ensemble of
$N$ atoms with a cavity initially prepared in a mesoscopic coherent state.
Using an appropriate mesoscopic approximation, we 
show that the strong atom/field interaction leads to an entangled  atom-field
state involving $N+1$ nearly coherent field components with different classical phases, 
generalizing the results obtained for one atom \cite{Gea:1991-1}. These coherent components are 
correlated with dipole atomic states, superpositions of the upper and lower states with equal weights. 
Thus, in the mesoscopic limit, the cavity field acts as a which-path detector for the atomic states 
interference. The periodic partial disentanglement of the atom-field system due to the transient 
overlap of field components is then closely linked to the complex pattern of quantum 
Rabi oscillation collapses and revivals observed in this regime. As in the single atom case, 
early quantum revivals can be induced by an echo sequence, realizing a time-reversal of 
the atom-field evolution  \cite{Morigi:2002-1}. The experimental observation of these 
effects would shed light on the deep links between entanglement and complementarity.

This complex phase-splitting was already predicted
by Knight and Shore \cite{Knight:1993-1}.
In the present paper, the introduction of an effective Hamiltonian
valid in the mesoscopic domain enables us to capture the main results within a
simple analytical model. This approach, originally pioneered by Klimov and Chumakov
\cite{Klimov:1995-1}, is also instrumental in
the discussion of dissipation in the system. 

Dissipation in the cavity turns the entangled atoms-field state into a statistical mixture, 
destroying Rabi oscillation revivals. In order to assess the experimental accessibility
of these mesoscopic quantum effects, we have analyzed quantitatively 
the influence of cavity dissipation on the
evolution of the atom-cavity entangled state. Using the physical insight
provided by the stochastic wave function 
approach \cite{Dalibard:1992-1} to the dissipative dynamics
 of the atoms + cavity system, an analytic formula for
the decoherence of the mesoscopic atoms + cavity state is
derived. It generalizes to the case of 
$N>1$ atoms the results previously obtained by Gea-Banacloche 
\cite{Banacloche:1993-1} in the $N=1$ case.  
We provide a functional expression for the decoherence coefficients
of the entangled atoms + cavity state which is valid even in the presence of an 
echo sequence used to induce an early revival of the quantum Rabi oscillation.
The functional form of these decoherence coefficients reflects the
cumulative construction of the imprint left by the strongly coupled atoms + cavity system in
the cavity environment. These physically illuminating expressions can be straightforwardly 
generalized to compute decoherence properties 
during a more complex protocol, such as the injection of 
another atomic ensemble in the cavity shortly after the first one in order 
to probe the cavity field. 

 \medskip
 
The organization of this paper is as follows: in section \ref{sec:effectiveH}, the
model for the resonantly coupled atoms + cavity system is presented and its
dynamics is studied in the absence of dissipation using a mesoscopic approximation
in the spirit of Gea-Banacloche \cite{Gea:1991-1}. In section \ref{sec:dissipative},
dissipation of the cavity is introduced and studied analytically using the
stochastic wave function approach. Section \ref{sec:results} presents 
numerical results obtained from quantum Monte-Carlo simulations. These results are used
to discuss experimentally accessible windows for the
observation of a mesoscopic entanglement between two or three atoms
in a microwave high quality cavity in the near future. We also comment on the
possibility of observing such mesoscopic effects within the context
of circuit-QED experiments performed with nanofabricated 
superconducting circuits \cite{Blais:2004-1}. The next generation of these 
experiments will involve several qubits coupled to a cavity. Therefore, 
it is very natural to address the question of entanglement 
between several qubits and the resonator for circuit-QED devices.
 
\section{Hamiltonian effective dynamics in the mesoscopic regime}
\label{sec:effectiveH}

\subsection{The Tavis-Cummings model}

In this paper, the resonant interaction between $N$ two-level atoms and an
electromagnetic mode in a cavity is considered. The cavity mode is modeled by a quantum 
harmonic oscillator which, in section \ref{sec:dissipative}, will be weakly coupled to an 
harmonic bath representing its environment. 

\medskip
 
Assuming that all atoms are symmetrically coupled to 
the mode, the atom-field system is conveniently described by the Tavis-Cummings model
\cite{Tavis:1968-1}, a spin $J=N/2$ 
generalization of the Jaynes-Cummings model \cite{Jaynes:1963-1}. 
The interaction between the atoms
and the electromagnetic mode is given by:
\begin{eqnarray}
\label{eq:Dicke}
H = \frac{\hbar g}{2}\sum_{i=1}^N\left(S_i^+\,a+S_{i}^-\,a^\dagger\right)
\end{eqnarray}
where $S_{i}^\pm$ denote the raising and lowering operators for the $i$th atom. The
energy scale associated with the interaction 
of one atom with the  mode is $\hbar g$.
Because of
the symmetric coupling, the evolution is restricted to the symmetric
subspace, invariant under atomic permutations, provided the 
initial state is also symmetric, a condition that we assume fulfilled from now on.
The atomic degree
of freedom is the spin $J=N/2$ representation for the 
collective $su(2)$ generators:
\begin{equation}
\label{eq:definition-SU2}
J^z= \sum_{i=1}^NS^z_{i}\, , \quad
J^\pm= \sum_{i=1}^NS^\pm_{i}\,.
\end{equation}
The interaction Hamiltonian can then be rewritten in terms of these operators
leading to the Tavis-Cummings (TC) model:
\begin{equation}
\label{eq:Dicke-SU2}
H_{TC} = \frac{\hbar g}{2}\,\left(J^+\,a+J^-\,a^\dagger\right)\ .
\end{equation}
Within this framework, the atomic ensemble behaves as a collective quantum object, a spin $J=N/2$ 
interacting with a quantum harmonic oscillator. A convenient basis in the atom $+$ cavity
Hilbert space is made up of tensor products of 
the atomic Dicke states $|J,m\rangle$, common eigenstates of $J^2$ and $J^z$, and
the Fock states $|n\rangle $ for the harmonic oscillator.
Note that the Hilbert space for this coupled system contains stable subspaces under time 
evolution which organize as follows: first, an infinity of $2J+1$
dimensional subspaces $\mathcal{H}_{n}$ ($n\geq 0$)
generated by the states $|J,J-l\rangle \otimes |n+l\rangle$ where
$l$ ranges from $0$ to $2J$. Then, a finite number of lower dimension
subspaces indexed by $-J\leq m<J-1$ 
generated by 
$|J,m-l\rangle \otimes |l\rangle $ where $0\leq l\leq J-m$.

In this paper, we focus on the mesoscopic regime in which the exchange of quanta
between the collective state of the $N$ atoms and the cavity mode does not significantly
alter the latter. Since the collective atomic spin can transfer at most $N$ photons to
the electromagnetic mode, this implies $\bar{n}\gg N$,
where $\bar{n}$ is the mean photon number in the cavity. 

\subsection{Mesoscopic entanglement involving one atom in a cavity}
\label{sec:Banacloche}

\subsubsection{Mesoscopic approximation for the atom + cavity evolution}

The quantum dynamics of a single atom interacting with a coherent state in a cavity has
been investigated by Gea-Banacloche \cite{Gea:1991-1} and independently by Buzek and Knight
\cite{Buzek:1992-1}.
The analysis by Gea-Banacloche is based on the
exact diagonalization of the Jaynes-Cummings Hamiltonian. It provides an approximate
solution for the Schrödinger equation with the initial condition
$|\psi_{\mathrm{at}}\rangle \otimes |\alpha\rangle$ where $|\psi_{\mathrm{at}}\rangle$ 
denotes the initial state
of the two-level atom and $|\alpha\rangle$ is a coherent state of the cavity field 
containing a mesoscopic number $\bar{n}=|\alpha|^2$ of photons ($\alpha=
\sqrt{\bar{n}}$). 

\medskip

The atom + field interaction is expected to create an entangled state. As noticed by
Knight and Shore, using an argument based on the Schmidt theorem \cite{Knight:1993-1}, a 
two-level atom cannot get entangled with more than two orthonormal states of the field. 
The Gea-Banacloche approximate solution precisely expresses 
the atom+field state $|\Psi(t)\rangle$ at time $t$
as a two-component entangled state. 
As discussed by Gea-Banacloche \cite{Gea:1991-1}, this approximation is accurate
for $t\ll \bar{n}/g$ which, for $\bar{n}\gg 1$, is large compared to the vacuum Rabi period 
$2\pi/g$. It leads to:
\begin{eqnarray}
\label{eq:entangledN1}
|\Psi(t)\rangle & = & A\,
 e^{-igt\sqrt{\bar{n}}/2}|D_{+}(t)\rangle \otimes |\psi_{+}(t)\rangle 
  \nonumber\\
 & + & B\,
 e^{igt\sqrt{\bar{n}}/2}|D_{-}(t)\rangle \otimes |\psi_{-}(t)\rangle 
\end{eqnarray}
where $A$ and $B$ characterize the initial atomic state ($|e\rangle$ in 
recent experiments \cite{Meunier:2005-1}).
The atomic dipole states $|D_{\pm}(t)\rangle$ are given by:
\begin{equation}
|D_{\pm}(t)\rangle=\frac{1}{\sqrt{2}}\left(\pm %e^{i\varphi/2}
\, e^{\mp igt/4\sqrt{\bar{n}}}|+\rangle +%e^{-i\varphi/2}
|-\rangle
\right)\,
\end{equation}
and the field states $|\psi_{\pm}(t)\rangle$ are:
\begin{equation}
|\psi_{\pm}(t)\rangle = e^{\pm \frac{igt\sqrt{\bar{n}}}{2}}\, 
e^{-\bar{n}/2} \sum_{k=0}^\infty
\frac{\alpha^k}{\sqrt{k!}}\,e^{\mp\frac{igt\sqrt{k}}{2}}|k\rangle\,.
\end{equation}
In the following, we will use the short hand denomination `Gea-Banacloche states' for these 
cavity states and their generalization to $N>1$.

\subsubsection{Discussion}

Gea-Banacloche has also shown 
that, for times short compared to
$g^{-1}\sqrt{\bar{n}}$, the state $|\psi_{\pm}(t)\rangle$ can be
approximated by a coherent state of parameter $\alpha_{\pm}(t)=e^{\mp igt/4\sqrt{\bar{n}}}
\alpha $. This result is obtained by expanding $\sqrt{k}$ at first order in 
$k-\bar{n}$ around $\sqrt{\bar{n}}$
leading to:
\begin{equation}
\label{eq:GB-coherent}
|\psi_{\pm}(t)\rangle \simeq 
e^{\pm igt\sqrt{\bar{n}}/4}\,|\alpha \,e^{\mp igt/4\sqrt{\bar{n}}}\rangle
\end{equation}
in the limit $t\ll \sqrt{\bar{n}}/g$.
Thus, the states $|\psi_{\pm}(t)\rangle$ mainly evolve at 
slow frequencies $\pm g/4\sqrt{\bar{n}}$.
We refer to \eqref{eq:GB-coherent} as the `coherent state approximation' for
Gea-Banacloche states and, if inserted in \eqref{eq:entangledN1}, 
as the coherent state approximation for
the atom + cavity system. As discussed in details in 
\cite{Gea:1991-1}, this approximation breaks down 
for $t\gtrsim\sqrt{\bar{n}}/g$ because the 
states $|\psi_{\pm}(t)\rangle$ undergo a slow phase spreading due to higher order terms in their 
expansion. They can no longer be considered as coherent. 
However, even if it breaks down before the mesoscopic approximation, the
coherent state approximation provides a nice intuitive and pictorial support
for visualizing the system's evolution. 

\medskip

\begin{figure}
\begin{center}
%% Picture
\begin{picture}(0,0)%
\epsfig{file=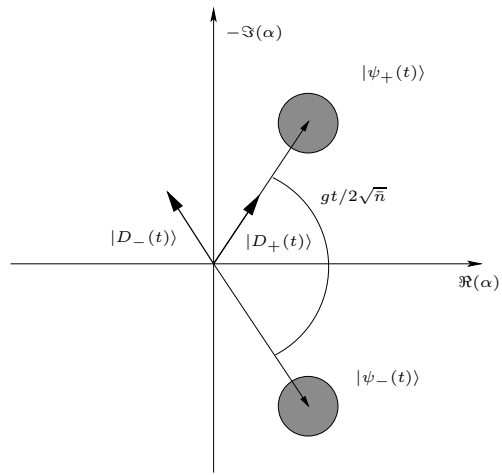}%
\end{picture}%
\setlength{\unitlength}{2072sp}%
\begingroup\makeatletter\ifx\SetFigFont\undefined%
\gdef\SetFigFont#1#2#3#4#5{%
  \reset@font\fontsize{#1}{#2pt}%
  \fontfamily{#3}\fontseries{#4}\fontshape{#5}%
  \selectfont}%
\fi\endgroup%
\begin{picture}(5667,5596)(2132,-6245)
\put(7501,-4018){\makebox(0,0)[lb]{\smash{{\SetFigFont{6}{7.2}{\familydefault}{\mddefault}{\updefault}{\color[rgb]{0,0,0}$\Re{(\alpha)}$}%
}}}}
\put(6338,-1501){\makebox(0,0)[lb]{\smash{{\SetFigFont{6}{7.2}{\familydefault}{\mddefault}{\updefault}{\color[rgb]{0,0,0}$|\psi_+(t)\rangle$}%
}}}}
\put(6271,-5146){\makebox(0,0)[lb]{\smash{{\SetFigFont{6}{7.2}{\familydefault}{\mddefault}{\updefault}{\color[rgb]{0,0,0}$|\psi_-(t)\rangle$}%
}}}}
\put(4715,-1018){\makebox(0,0)[lb]{\smash{{\SetFigFont{6}{7.2}{\familydefault}{\mddefault}{\updefault}{\color[rgb]{0,0,0}$-\Im{(\alpha)}$}%
}}}}
\put(4951,-3526){\makebox(0,0)[lb]{\smash{{\SetFigFont{6}{7.2}{\familydefault}{\mddefault}{\updefault}{\color[rgb]{0,0,0}$|D_+(t)\rangle$}%
}}}}
\put(5851,-2986){\makebox(0,0)[lb]{\smash{{\SetFigFont{6}{7.2}{\familydefault}{\mddefault}{\updefault}{\color[rgb]{0,0,0}$gt/2\sqrt{\bar{n}}$}%
}}}}
\put(3331,-3481){\makebox(0,0)[lb]{\smash{{\SetFigFont{6}{7.2}{\familydefault}{\mddefault}{\updefault}{\color[rgb]{0,0,0}$|D_-(t)\rangle$}%
}}}}
\end{picture}%
%%%
\end{center}
\caption{\label{fig:1}
Schematic evolution of the entangled state for one atom and a mesoscopic coherent
state in a cavity for real positive $\alpha$. The atomic dipole states are 
represented as arrows. The field coherent states are represented as an uncertainty disk
at the tip of the classical amplitude. Each component 
$|D_{\pm}(t)\rangle \otimes |\psi_{\pm}(t)\rangle$ 
of the superposition involves an atomic polarization 
and a field state slowly rotating in the phase plane 
at velocities $\pm g/4\sqrt{\bar{n}}$.}
\end{figure}

With this image in mind, it is useful to draw on the same diagram
the motion of the average atomic polarization $\overrightarrow{d_{\pm}}(t)
=\langle D_{\pm}(t)|\overrightarrow{\sigma}|D_{\pm}(t)\rangle$ in
the equatorial plane of the Bloch sphere
and the motion of $\alpha_{\pm}^*(t)$ in  the Fresnel plane.
The result is
depicted on figure \ref{fig:1} for $\varphi=0$: the corresponding vectors 
rotate at angular velocity $\pm g/4\sqrt{\bar{n}}$, small compared to
the classical Rabi frequency $g\sqrt{\bar{n}}$. 

In the limit $\bar{n}\rightarrow \infty$ and $gt\ll 1$, both states $|\psi_{\pm}(t)\rangle$ are
close to $|\alpha\rangle$ for $t\ll g^{-1}$, meaning that the cavity
mode is barely affected by the atoms. In this regime, the cavity state factors out and 
the atomic polarizations $|D_{\pm}(t)\rangle $ coincide with 
the atomic spin-1/2 eigenstates along the $x$ direction. These atomic states interfere
resulting in the classical Rabi oscillation phenomenon. Remember
that the period of classical Rabi oscillations is of the order $g^{-1}/\sqrt{\bar{n}}$.
Therefore, in the classical limit $\bar{n}\rightarrow +\infty$, more and more oscillations
take place before the motion of Gea-Banacloche states in the phase plane has any measurable consequence.

\medskip

In the mesoscopic regime (fixed $\bar{n}\gg 1$), the state of the cavity is altered by the atom. 
Interferences between atomic polarizations $|D_{\pm}(t)\rangle$
can only be observed when $|\psi_{+}(t)\rangle $ and $|\psi_{-}(t)\rangle$
overlap. 
As explained above, at very short times, these states are still close
to the initial coherent state $|\alpha\rangle$. When the phase separation between
$|\psi_{+}(t)\rangle $ and $|\psi_{-}(t)\rangle$ due to their slow rotation in phase space is 
larger than the quantum phase fluctuations in these coherent 
components ($gt/2\sqrt{\bar{n}}\sim 1/\sqrt{\bar{n}}$), the cavity field 
behaves as a {\sl bona fide} ``path detector" for the atomic polarizations and Rabi oscillations disappear. 
The Rabi oscillation collapses after a time
of the order of the vacuum Rabi oscillation, after $\sqrt{\bar{n}}$ classical oscillations.

\begin{figure}
\includegraphics[width=9cm]{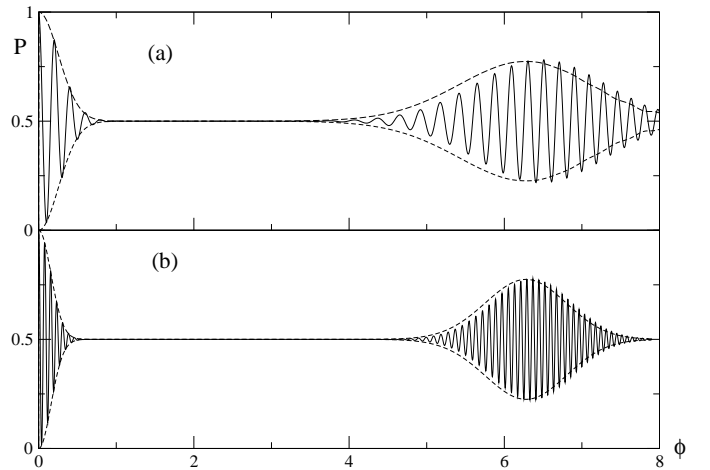}
\caption{\label{fig:4}
Spontaneous revivals of the Rabi oscillation signal for one atom with
initial condition $(|e\rangle=|m=1/2\rangle )\otimes |\alpha\rangle $ and a coherent state
with mean photon number (a) $\bar{n}=15$ and (b) $\bar{n}=40$. 
The solid line shows the probability $P(\phi)$ for finding the atom in $|e\rangle$, 
as a function of the dimensionless time $\phi=gt/2\sqrt{\overline{n}}$, computed
using numerical integration. The dashed curves show the 
upper and lower envelopes predicted by our mesoscopic approximation.}
\end{figure}

The Rabi oscillation signal
reappears when $|\psi_{+}(t)\rangle$ and $|\psi_{-}(t)\rangle$ overlap again. This
happens for $gt/2\sqrt{\bar{n}}\simeq 2\pi$. During this overlap, 
the disentanglement of the atom + cavity state
erases the information stored in the cavity about the path followed by the atomic
degrees of freedom. This `quantum eraser situation' leads to a revival of Rabi oscillations. 
Rabi oscillation
revivals in the mesoscopic regime are thus a direct application of the complementarity 
concept \cite{Rauschenbeutel:2001-1}. Figure \ref{fig:4}  shows, 
as a function of the dimensionless time $\phi=gt/2\sqrt{\overline{n}}$,
the first spontaneous revival of the Rabi oscillation
signal obtained by numerical integration of the Schrödinger equation for one
atom initially in the excited state and coherent states of 14 and 40 photons in average.

\subsubsection{The echo protocol}

The echo protocol proposed by Morigi \cite{Morigi:2002-1} aims at testing the coherence
of the atom + cavity state by a time reversal operation. 
A percussional echo pulse is applied to the atom at time
$t_{\pi}$. It corresponds to the unitary operator 
$U_{\pi}=i\,\sigma^z$. The evolution for the atom + cavity 
system up to time $t\geq t_{\pi}$ is then given by
\begin{equation}
\label{eq:echo:evolution}
U(t)=e^{-i(t-t_{\pi})H/\hbar}\ldotp U_{\pi}\ldotp e^{-it_{\pi}H/\hbar}
\end{equation}
where $H$ is the Jaynes-Cummings Hamiltonian.
Using $U_{\pi}^2=1$ and $U_{\pi}HU_{\pi}=-H$, we get:
\begin{equation}
\label{eq:echo:evolution:result}
U(t)=U_{\pi}\ldotp e^{-i(t_{\pi}-t)H/\hbar}\ldotp e^{-it_{\pi}H/\hbar}\,.
\end{equation}
Therefore, right after the echo pulse, the Gea-Banacloche states reverse their evolution and
recombine at $2t_{\pi}$ leading to an induced Rabi oscillation revival. This
induced revival may occur at much shorter time than the `spontaneous' revival, making 
its experimental observation much easier, as shown recently \cite{Meunier:2005-1}. Moreover, in 
the absence of decoherence, the induced revival should occur with unit contrast. The influence of decoherence could thus be, in principle, directly assessed from the measurement of the 
induced revival contrast.

\subsubsection{Towards atomic ensembles}

In this paper, we are interested in studying the resonant interaction of an atomic ensemble containing 
$N>1$ atoms with a mesoscopic field in a cavity. Invoking again Schmidt theorem \cite{Knight:1993-1}, 
we expect this resonant interaction to create an
entangled state with $2J+1=N+1$ orthonormal components. As in the one-atom case,
partial disentanglement of this state will lead to spontaneous revivals of Rabi oscillations. 

\medskip

An analytic diagonalization of the 
Tavis-Cummings Hamiltonian can be obtained for $N=1,2,3$ but not
for greater values of $N$. Moreover, as will be clear from forthcoming sections,
the analytical diagonalization for these values of $N$ does not enlighten the
dynamics of the system. In particular, for $N>1$, the explicit expressions of
exact eigenstates in the resonant Tavis-Cummings Hamiltonian depend on $N$. 
This direct approach thus cannot be 
used as a convenient starting point for an approximate
solution of the Schr\"odinger equation for $N>1$. 

Our approach, developped in the next section, relies on an 
effective Hamiltonian which, in the mesoscopic
domain, provides an excellent approximation to the Tavis-Cummings Hamiltonian. 
It provides a unified vision of the dynamics for all values of $N$ and as such,
it is a good starting point for analyzing the dynamics in the mesoscopic domain. 
As we shall see, in this framework, the dynamics of the resonant Tavis-Cummings model
can then be described in the spirit of the Gea-Banacloche approach.
%With this effective Hamiltonian, an approximate solution to the 
%Schrödinger equation with initial
%condition $|J,m_{0}\rangle \otimes |\alpha\rangle$ will be found, in the 
%spirit of the Gea-Banacloche approach. 

\subsection{Effective dynamics in the mesoscopic domain}

\subsubsection{Effective Hamiltonian}

First of all, let us remark that any
initial state of the form $|J,m_{0}\rangle \otimes |\alpha\rangle$ in the mesoscopic
domain mainly spreads over $(2J+1)$-dimensional stable subspaces $\mathcal{H}_{n}$ for
values of $n$ around $\bar{n}$. The core of our approach is to replace the 
Tavis-Cummings Hamiltonian \eqref{eq:Dicke-SU2} acting on subspaces $\mathcal{H}_{n}$ by an
effective Hamiltonian in which the $n$ dependence factors out. It appears that 
the main $n$ dependence of \eqref{eq:Dicke-SU2} scales 
as $\sqrt{n}$ for large values of $n$. As in the $N=1$ case, this non-linearity
leads to the collapse of the Rabi oscillations and the discrete character of the spectrum 
leads to spontaneous revivals.

\medskip

\begin{figure}
\begin{center}
\begin{picture}(0,0)%
\epsfig{file=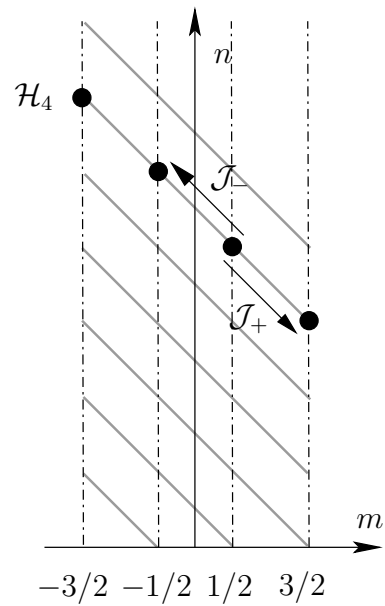}%
\end{picture}%
\setlength{\unitlength}{4144sp}%
\begingroup\makeatletter\ifx\SetFigFont\undefined%
\gdef\SetFigFont#1#2#3#4#5{%
  \reset@font\fontsize{#1}{#2pt}%
  \fontfamily{#3}\fontseries{#4}\fontshape{#5}%
  \selectfont}%
\fi\endgroup%
\begin{picture}(2371,3594)(4321,-5803)
\put(5514,-2559){\makebox(0,0)[lb]{\smash{{\SetFigFont{12}{14.4}{\familydefault}{\mddefault}{\updefault}{\color[rgb]{0,0,0}$n$}%
}}}}
\put(5492,-3302){\makebox(0,0)[lb]{\smash{{\SetFigFont{12}{14.4}{\familydefault}{\mddefault}{\updefault}{\color[rgb]{0,0,0}$\mathcal{J}_-$}%
}}}}
\put(4321,-2813){\makebox(0,0)[lb]{\smash{{\SetFigFont{12}{14.4}{\familydefault}{\mddefault}{\updefault}{\color[rgb]{0,0,0}$\mathcal{H}_4$}%
}}}}
\put(5604,-4134){\makebox(0,0)[lb]{\smash{{\SetFigFont{12}{14.4}{\familydefault}{\mddefault}{\updefault}{\color[rgb]{0,0,0}$\mathcal{J}_+$}%
}}}}
\put(6369,-5348){\makebox(0,0)[lb]{\smash{{\SetFigFont{12}{14.4}{\familydefault}{\mddefault}{\updefault}{\color[rgb]{0,0,0}$m$}%
}}}}
\put(5469,-5753){\makebox(0,0)[lb]{\smash{{\SetFigFont{12}{14.4}{\familydefault}{\mddefault}{\updefault}{\color[rgb]{0,0,0}$1/2$}%
}}}}
\put(4958,-5754){\makebox(0,0)[lb]{\smash{{\SetFigFont{12}{14.4}{\familydefault}{\mddefault}{\updefault}{\color[rgb]{0,0,0}$-1/2$}%
}}}}
\put(5904,-5746){\makebox(0,0)[lb]{\smash{{\SetFigFont{12}{14.4}{\familydefault}{\mddefault}{\updefault}{\color[rgb]{0,0,0}$3/2$}%
}}}}
\put(4457,-5753){\makebox(0,0)[lb]{\smash{{\SetFigFont{12}{14.4}{\familydefault}{\mddefault}{\updefault}{\color[rgb]{0,0,0}$-3/2$}%
}}}}
\end{picture}%
\end{center}
\caption{\label{fig:0}
Schematic view of the Hilbert space for $N=3$ atoms. The oblique grey lines represent the
stable subspaces. The action of $\mathcal{J}_{\pm}$ on $\mathcal{H}_{3}$ is
depicted.}
\end{figure}

In order to describe our ansatz for the effective Hamiltonian, it  is convenient to
remark that each subspace $\mathcal{H}_{n}$ 
can be turned into a spin-$J$ representation of $su(2)$. Let us introduce new
operators $\mathcal{J}^\pm$ and $\mathcal{J}^z$. With the notation
$|Z^{(n)}_{m}\rangle = |J,m\rangle \otimes |n+J-m\rangle$, these new operators
simply act on these states in the same way as standard $su(2)$ generators act on the $|J,m\rangle$ states:
$\mathcal{J}^\pm|Z^{(n)}_{m}\rangle = \sqrt{J(J+1)-m(m\pm 1)}\,|Z^{(n)}_{m\pm 1}\rangle$
(see fig. \ref{fig:0}). The operators $a\,J^+$ and $a^\dagger\,J^-$ then act on the
states $|Z^{(n)}_{m}\rangle $ as:
\begin{eqnarray}
a\,J^+ |Z^{(n)}_{m}\rangle & = & \sqrt{n+J-m}\,\mathcal{J}^+ |Z^{(n)}_{m}\rangle \\
a^\dagger\,J^- |Z^{(n)}_{m}\rangle & = & \sqrt{n+J-m+1}\,
\mathcal{J}^- |Z^{(n)}_{m}\rangle\,. 
\end{eqnarray}
We then note that computing the evolution of a state $|J,m\rangle\otimes |\alpha\rangle$
in the mesoscopic regime
requires considering values of $n$ close to $\bar{n}\gg N$. The variation of $\sqrt{n+k}$ for
$0\leq k\leq N+1$ is small for $n\gg N$ (of the order of $N/\sqrt{\bar{n}}$).
We thus drop the $m$-dependence of  $\sqrt{n+J-m}$
and $\sqrt{n+J-m+1}$ by replacing them by $\sqrt{n+c}$, where $0\leq c\leq N+1$ is a constant 
to be discussed in the next paragraph.
This leads to an effective Hamiltonian of the form:
\begin{equation}
\label{eq:Heffective}
H_{\mathrm{eff}}^{(n)}=
\frac{ \hbar g}{2} \,\sqrt{n+c}\,
(\mathcal{J}^++\mathcal{J}^-)=\hbar g\sqrt{n+c}\,\,\mathcal{J}^x\,.
\end{equation}
This Hamiltonian, already derived by Klimov and Chumakov
\cite{Klimov:1995-1} and used to study the squeezing of light by an atomic ensemble 
\cite{Retamal:1997-1}, shares some features with the
expected classical dynamics, driven by an effective field along $x$. Here however,
photon emission and absorption are taken into account through the fact that 
$\mathcal{J}^\pm$ changes the photon number (see fig. \ref{fig:0}).
At fixed $n$, the eigenvalues of this effective Hamiltonian are equally spaced, as 
predicted by \cite{Tavis:1968-1} for the Tavis-Cummings
Hamiltonian in the large $n$ limit. All the $n$ dependence of this effective Hamiltonian is 
contained in the $\sqrt{n+c}$ factor.

Of course, there is an ambiguity in the choice of $0\leq c\leq N+1$ 
but we shall see that {\it (i)} for $N=1$
choosing $c=1$ reproduces the results of section \ref{sec:Banacloche} and {\it (ii)} 
for $N>1$, changing $c$ only affects the rapidly oscillating part of the Rabi oscillation 
signal. It does not change its envelope which is precisely
the information we hope to extract from the effective Hamiltonian. 

\subsubsection{The mesoscopic approximation}

Using this effective Hamiltonian, it is possible to
study the evolution of a state $|\Psi_{m}^X\rangle=
|J,m\rangle_{x}\otimes |\alpha\rangle$
where $J^x|J,m\rangle_{x}=m\,|J,m\rangle_{x}$. 
An approximate solution for the Schrödinger equation 
shows that this state remains factorized (see appendix \ref{sec:appendixA} for details):
\begin{equation}
\label{eq:entangled-state}
|\Psi_{m}^X(t)\rangle = e^{-imgt|\alpha|}\,|D_{m}(t)\rangle \otimes
|\psi_{m}(t)\rangle
\end{equation}
where the state of the electromagnetic mode is of the form:
\begin{equation}
\label{eq:GBstate}
|\psi_{m}(t)\rangle = e^{imgt\sqrt{\bar{n}}}\, 
e^{-\bar{n}/2} \sum_{k=0}^\infty
\frac{\alpha^k}{\sqrt{k!}}\,e^{-imgt\sqrt{k}}|k\rangle\,,
\end{equation}
which we call, as above, a Gea-Banacloche state \cite{Gea:1991-1}.
The atomic polarizations generalize the ones found by Gea-Banacloche in the spin $1/2$
case: 
\begin{equation}
\label{eq:atomic-polarization}
|D_{m}(t)\rangle=\sum_{m'=-J}^J e^{-igmt(c-J+m')/2\sqrt{\bar{n}}}(R^{-1})_{m,m'}\,
|J,m'\rangle\,.
\end{equation}
where $R$ denotes the rotation matrix 
$R_{m,m'}=\langle J,m'|e^{i\pi J^y/2}|J,m\rangle$. Note the presence of the
classical Rabi frequencies $mg|\alpha|$ corresponding to the quantum beat between spin eigenstates along 
the $x$ direction of the effective classical field.
The average angular momentum
$\overrightarrow{d_{m}}(t)=\langle D_{m}(t)|\overrightarrow{J}|D_{m}(t)\rangle$
slowly rotates in the equatorial plane of the Bloch sphere
at angular velocity $gm/2\sqrt{\bar{n}}$. 
The parameter $c$ appears in these atomic polarizations only and,
for $N=1$, the Gea-Banacloche results are exactly recovered for $c=1$.

Starting from state $|\Psi(0)\rangle=|J,m\rangle\otimes|\alpha\rangle$, 
an entangled state with $N+1$ components is obtained:
\begin{equation}
\label{eq:entangled-state}
|\Psi(t)\rangle = \sum_{m=-J}^JR_{m_{0},m}\,e^{-imgt|\alpha|}
\,|D_{m}(t)\rangle \otimes
|\psi_{m}(t)\rangle\,.
\end{equation}
As in the one-atom case, the entangled state \eqref{eq:entangled-state} 
can be viewed as the result of
the ideal measurement of the spin by the mesoscopic field in the cavity. 

\subsubsection{The coherent state approximation}

As in the single-atom case, 
the state $|\psi_{m}(t)\rangle$ can be
approximated by a coherent state of complex amplitude $\alpha_{m}(t)=e^{-imgt/2\sqrt{\bar{n}}}
\alpha $. This approximation holds in the limit
$t\ll (g|m|)^{-1}\sqrt{\bar{n}}$. At longer times, typically $\sqrt{\bar{n}}/g|m|$, the field state
gets deformed as the $|\psi_{\pm}\rangle$ states in the
$N=1$ case. 

With this image in mind, it is useful to draw on the same diagram
the motion of the average atomic polarization $\overrightarrow{d_{m}}(t)$ in
the equatorial plane of the Bloch sphere 
and the motion of $\alpha_{m}(t)$ in  the Fresnel plane, generalizing the phase 
space representation used above. The main difference with the $N=1$ case
is the appearance of $N+1$ frequencies and field states instead of two. Here also, the phase
of the coherent state plays the role of a pointer measuring the angular momentum of the
collective spin along the $x$ direction in the $\bar{n}\gg 1$ 
limit. Larger angular momenta lead to larger angular velocities.
As we shall see now, this complex atoms + cavity entangled state
leads to a rich pattern of spontaneous revivals of Rabi oscillations.

\subsection{Partial revivals of Rabi oscillations}

\subsubsection{General picture}

Rabi oscillations of the atomic populations provide a nice way to probe the 
degree of entanglement of the atom + cavity state. 

In the classical limit 
($\bar{n}\rightarrow \infty$), the electromagnetic field state factors out and 
quantum interferences between the various atomic 
polarizations $|D_{m}(t)\rangle$ can be observed. They are the
Rabi oscillations for the quantum spin $J$ in a transverse classical field.

In the mesoscopic limit, the electromagnetic
mode is altered by the atom. Interferences between atomic polarizations $|D_{m}(t)\rangle$
can only be observed when the corresponding field states $|\psi_{m}(t)\rangle $ 
overlap. At very short times, the various components  $|\psi_{m}(t)\rangle $  are still close
to the initial coherent state $|\alpha\rangle$ and Rabi oscillations  show up. When the
various Gea-Banacloche states split apart, the electromagnetic field becomes a good ``path
detector" for the atomic polarizations and the Rabi oscillations collapse. This is again a complementarity 
effect, the field storing a which-path information about the interfering atomic states. 
Rabi oscillations
reappear when this which-path information is, at least partially, erased, {\it i.e.} 
when some of the Gea-Banacloche states overlap again.
In the $N>1$ case, the atoms + cavity state is a superposition of $N+1$ factorized 
components rotating at different velocities and
a rich spontaneous revival pattern is expected. 

\medskip

The angular velocity of
the Gea-Banacloche states $|\psi_{m}(t)\rangle$
in the Fresnel plane is here $-gm/2\sqrt{\bar{n}}$, suggesting to associate
with each Rabi oscillation revival a non-empty 
subset $\mathcal{E}$ of $\{1,\ldots ,2J\}$ which is the list of absolute values of differences
of the indices $m$ of those Gea-Banacloche states that overlap during the revival under consideration. 
Given such a subset $\mathcal{E}$, the revival is built from contributions of pairs
of states $|\psi_{m_{+}}(t)\rangle$ and $|\psi_{m_{-}}(t)\rangle$ such that 
$|m_{+}-m_{-}|\in \mathcal{E}$. 
Note that, when a pair 
$|\psi_{m_{+}}(t)\rangle$ and $|\psi_{m_{-}}(t)\rangle$ overlaps, all pairs that
have a positive or negative integer multiple value of $m_{-}-m_{+}$ also overlap. Therefore
if $q$ belongs to such a subset, its multiples
also do. This is the only constraint on the subsets $\mathcal{E}$. The Rabi oscillation
revivals are therefore classified by the greatest common divisor 
$\mathrm{gcd}(\mathcal{E})$  of the elements 
of $\mathcal{E}$.
The first time of
occurrence of the spontaneous revival associated with $\mathcal{E}$
is $t$ such that $gt/2\sqrt{\bar{n}}=2\pi / \mathrm{gcd}(\mathcal{E})$. Replicas
of this revival will occur at integer multiples of this fundamental time. Note that there are
$N+1-\mathrm{gcd}(\mathcal{E})$ pairs of Gea-Banacloche states that verify
$|m_{+}-m_{-}|=q$.
Remember that
$\phi=gt/2\sqrt{\bar{n}}$ is 
the dimensionless time which characterizes the slow motion of Gea-Banacloche states. 
In general, the
contrast of replicas will be reduced because of the spreading of the Gea-Banacloche state (especially
if they occur after $gt/2\sqrt{\bar{n}}\geq2\pi$).
For all values of $N$, the set $\{1,\ldots ,N\}$ corresponds to a complete revival involving the
recombination of all Gea-Banacloche states at $\phi=2\pi$. 

\begin{figure}
\includegraphics[width=9cm]{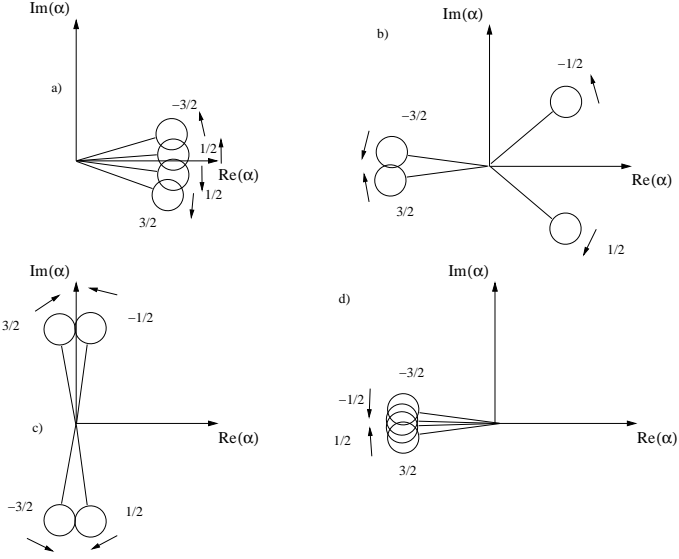}
\caption{\label{fig:3}
Position of the Gea-Banacloche states at the times of first occurrence of spontaneous 
Rabi oscillation revivals for $N=3$ ($J=3/2$): (a) Collapse of the Rabi oscillations when the
various components split apart (b) First spontaneous revival for $\phi=2\pi/3$ (c) Second spontaneous 
revival  for $\phi=\pi$ (d) Complete spontaneous revival involving all atomic 
polarizations for $\phi=2\pi$. 
}
\end{figure}

As an example, let us consider the case of three atoms. 
The corresponding Rabi revival patterns are depicted on Figure \ref{fig:3}.
The first revival is obtained 
when $m=\pm 3/2$ states overlap for $\phi=2\pi/3$
(associated subset $\mathcal{E}=\{3\}$).
It is partial (contrast is lower than one) since only two atomic polarizations 
take part in it. The next revival appears for $\phi=\pi$ when $m=3/2$ \& 
$m=-1/2$ and  $m=-3/2$ \& $m=1/2$ overlap separately ($\mathcal{E}=\{2\}$). 
At $\phi=4\pi/3$, $m=3/2$ and $m=-3/2$ recombine again leading to a partial revival 
which is a replica of the first one ($\mathcal{E}=\{3\}$).
Finally the complete revival
involving quantum interferences between all four atomic polarizations takes place at
$\phi=2\pi$ ($\mathcal{E}=\{1,2,3\}$). Note that the first example of a non-trivial revival involving
several slow frequencies before the complete revival occurs for $N=4$ at $\phi=\pi$ 
($\mathcal{E}=\{2,4\}$).

\subsubsection{Rabi oscillation envelopes}
\label{sec:effectiveH:enveloppes}

Preparing $J+m_{0}$ atoms in the excited state,
the probability of detecting $J+m$ ones in the excited state and $J-m$ in
the ground state is given by ($q=m_{+}-m_{-}$):
\begin{equation}
P_{m}(t)=\sum_{m_+,m_-}
e^{-igqt\sqrt{\bar{n}}/2}\mathcal{P}^{(m_{0},m)}_{m_{+},m_{-}}(t)\,
\mathcal{R}_{m_+,m_-}(t)
\end{equation}
where $m_{\pm}$ run from $-J$ to $J$ and
\begin{equation}
\label{eq:defP}
\mathcal{P}^{(m_{0},m)}_{m_{+},m_{-}}(t)=
R_{m_0,m_+}R_{m_0,m_-}^*\,
\langle J,m|D_{m_+}(t)\rangle \langle D_{m_-}(t)|J,m\rangle \,
\end{equation}
contains the matrix elements of atomic polarizations.
The scalar products $\langle J,m|D_{m_\pm}(t)\rangle$
can also be expressed in terms of the rotation matrices (see appendix 
\ref{sec:appendixSU2} for explicit expressions):
\begin{equation}
\langle J,m|D_{m_\pm}(t)\rangle =
(R^{-1})_{m_{\pm},m}\,e^{-ig(c-J+m)m_{\pm}t/2\sqrt{\bar{n}}}\,.
\end{equation}
Within our effective Hamiltonian approximation, the time dependence of the 
atomic polarization factor $\mathcal{P}^{(m_{0},m)}_{m_-,m_+}$ is a phase.
Modulation factors for the revivals come from the overlaps of cavity mode states:
\begin{equation}
\label{eq:defR}
\mathcal{R}_{m_+,m_-}(t)= 
\langle \psi_{m_-}(t)|\psi_{m_+}(t) \rangle \,.
\end{equation}
This expression only depends on $q=m_{+}-m_{-}$. Finally, the Rabi oscillation signal
is
\begin{equation}
\label{eq:Rabi:splittedfreq}
P_{m}(t)=\sum_{q=-2J}^{2J}
\mathcal{R}_{q}(t)\,\mathcal{A}^{(m_{0},m)}_{q}e^{-i\frac{gqtc}{2\sqrt{\bar{n}}}}
\,e^{-igqt\sqrt{\bar{n}}}
\end{equation}
where 
$$\mathcal{A}^{(m_0,m)}_{q}=\sum_{m_--m_+=q}R_{m_0,m_+}
R_{m_0,m_-}^*R^{-1}_{m,m_+}(R^{-1}_{m,m_-})^*.$$
This expression
separates the rapid frequencies $gq\sqrt{\bar{n}}$ from the mesoscopic slow frequencies
$gq/2\sqrt{\bar{n}}$. The Rabi signal then consists into a rapidly oscillating signal slowly 
modulated in amplitude and phase. Expression \eqref{eq:Rabi:splittedfreq}
can be used to find an approximate analytic expression for
the upper and lower envelopes of the signal. 

Let us illustrate this point on the signal
obtained for $m_{0}=m=J$ (denoting $\mathcal{A}^{(J,J)}_{q}=\mathcal{A}_{q}$)
which is plotted in the forthcoming figures. 

\medskip

Outside the spontaneous revivals, the contribution of the $q\neq 0$ terms in \eqref{eq:Rabi:splittedfreq}
vanishes. The base line of the Rabi oscillation signal is thus $\mathcal{A}_{0}$. Obtaining
the envelopes is trivial for $N=1$ since there is exactly one value of $q$ involved: $q_{r}=1$.
The slowly
varying phase $e^{-igqtc/2\sqrt{\bar{n}}}$ simply shifts the rapid oscillation without
changing its envelope. This analysis is correct also for $N>1$ in the case of a revival involving
exactly one frequency corresponding to $q_{r}\in \{1,\ldots N\}$ ($q_{r}=1$ for $N=1$).
For these revivals only, the upper and lower envelopes $P_{+}$ and $P_{-}$ take the form:
\begin{equation}
\label{eq:enveloppes:N=1}
P_{\pm}(t) = \mathcal{A}_{0}\pm
|\mathcal{R}_{q_{r}}(t)\,\mathcal{A}_{q_{r}}|\,.
\end{equation} 
Thus, simple revivals are symmetric with respect to the flat signal $\mathcal{A}_{0}$
and involve only one rapid frequency $gq_{r}\sqrt{\bar{n}}$.

The analysis turns out to be more involved when several frequencies are involved. A first
example of this situation is the initial collapse of Rabi oscillation ($t\lesssim 2\pi/g$)
for $N>1$. Nevertheless, the envelopes can be obtained using the exact expression for
classical Rabi oscillation in a field of amplitude
$\sqrt{\bar{n}}$: in this limit, the probability for detecting all atoms
in the excited state is given by $P_{c}(t)=\cos^{2N}(gt\sqrt{\bar{n}}/2)$. 
Its maxima occur at times $2\pi k/g\sqrt{\bar{n}}$
for integer values of $k$ and its minima occur for half integer values of $k$. Substituting these values
in \eqref{eq:Rabi:splittedfreq} for large
values of $\bar{n}$ provides the values of the rapidly oscillating term to be used
to fit the maxima (upper envelope) and the minima (lower envelope). This leads
to:
\begin{eqnarray}
\label{eq:enveloppe:N>1:upper}
P_{+}(t) & = & \mathcal{A}_{0}+\sum_{q\neq 0}
|\mathcal{R}_{q}(t)\,\mathcal{A}_{q}|\\
\label{eq:enveloppe:N>1:lower}
P_{-}(t) & = & \mathcal{A}_{0}+\sum_{q\neq 0}(-1)^q
|\mathcal{R}_{q}(t)\,\mathcal{A}_{q}|\,.
\end{eqnarray}
Note that for $N>1$, the envelope is not symmetric with respect to the flat signal 
$\mathcal{A}_{0}$.

Let us now turn to the complete revival which 
takes places around $t_{R}=4\pi \sqrt{\bar{n}}/g$. Near this
revival, the Rabi oscillation signal takes the form ($t=t_{R}+\tau$):
\begin{equation}
\label{eq:Rabi:main-resurgence}
P(t)=\mathcal{A}_{0}+\sum_{q\neq 0}
\mathcal{A}_{q}\mathcal{R}_q(t_{R}+\tau)\,e^{-i\frac{qg\tau}{2\sqrt{\bar{n}}}}\,
e^{-igq(\tau\sqrt{\bar{n}}+2\pi(c+2\bar{n}))}\,.
\end{equation}
We first note that the rapidly oscillating phases are shifted
in time by $2\pi(c+2\bar{n})/\sqrt{\bar{n}}$. This time shift does not
affect the low frequency modulating terms $\mathcal{A}_{q}\mathcal{R}_{q}(t)$.
Within the coherent state approximation, the overlap factors 
$\mathcal{R}_q(t_{R}+\tau)$ can be approximated by $1$ for $|\tau|\lesssim 2\pi/g$.
This means that, in the classical limit and within the coherent state approximation, 
close to the complete revival, the Rabi oscillation signal
has the same fast oscillations than near $t=0$. 
This
suggests to use \eqref{eq:enveloppe:N>1:upper} and \eqref{eq:enveloppe:N>1:lower} as
upper and lower envelopes. Because the Gea-Banacloche states are getting deformed 
over a time scale $t_{R}$, these expressions only provide an approximation to the real envelopes
of the theoretical signal \eqref{eq:Rabi:main-resurgence}. This approximation assumes
that the overlap factors $\mathcal{R}_{q}(t_{R})$ do not depend on $q$. Because 
$|\mathcal{R}_{q}(t_{R})-1|$ goes to zero as $1/\bar{n}$ in the large $\bar{n}$
limit, the accuracy  of \eqref{eq:enveloppe:N>1:upper} and 
\eqref{eq:enveloppe:N>1:lower} as approximate upper and
lower envelopes for the main spontaneous resurgence increases with 
increasing $\bar{n}$. 
%The general case is more involved and no general 
%simple prescription can be given for
%the upper and lower enveloppes. 

To summarize, \eqref{eq:enveloppe:N>1:upper} describes
the upper envelope of all revivals. The lower envelope is described by
$P_-(t)$ in eq. \eqref{eq:enveloppes:N=1} for revivals involving only one 
frequency such as the ones occuring at $gt/2\sqrt{\bar{n}}=2\pi/q$ where $
\left[N/2\right]< q\leq N$ and the lower envelope of the main revival
which occurs at $gt/2\sqrt{\bar{n}}=2\pi$ is described by \eqref{eq:enveloppe:N>1:lower}.
In practice, only the lowest values of $N$ ($N\leq 3$) may be easily reachable
in the Rydberg atoms experiments and therefore, the only revival involving more than one value
of $|m_{+}-m_{-}|$ is the complete one. 

\subsubsection{Numerical results}

All numerical results in this paper 
are presented in terms of the dimensionless time $\phi=gt/2\sqrt{\bar{n}}$ associated
with the slow evolution induced by the atom + field interaction.

\medskip

Let start by considering the case of $N=1$ atom. Figure \ref{fig:4} shows
the comparison between the analytic envelopes \eqref{eq:enveloppes:N=1} 
and an exact numerical solution
of the Schrödinger equation for different values of $\bar{n}$. As expected, the mesoscopic
approximation becomes better and better as $\bar{n}$ increases. In this case,
the upper and lower envelopes of the mesoscopic approximation signal are obtained
by setting $q_{r}=1$ in \eqref{eq:enveloppes:N=1}.

\medskip

\begin{figure}
\includegraphics[width=9cm]{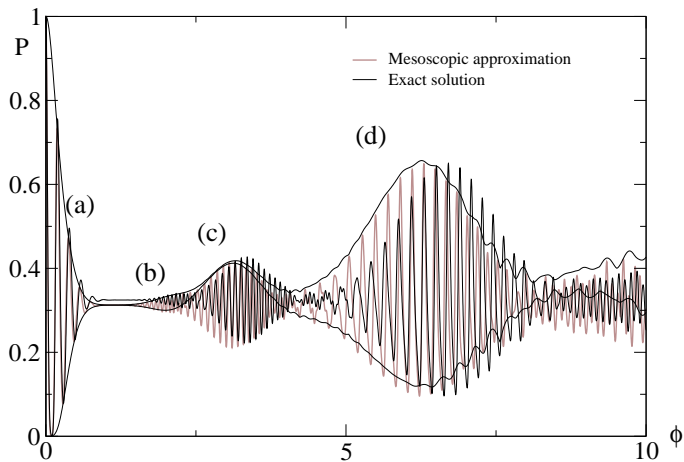}
\caption{\label{fig:5}
Spontaneous revivals of the Rabi oscillation signal $P=P_{m=J}$ for $N=3$ atoms ($J=3/2$), as 
a function of the dimensionless time $\phi=gt/2\sqrt{\bar{n}}$. The
initial state $|m=3/2\rangle \otimes |\alpha\rangle $ with an average photon number $\bar{n}=|\alpha|^2=15$. The signal is
computed using the mesoscopic approximation (grey line), its upper and lower envelopes (plain 
lines) and the exact solution (plain line with rapid oscillations)
in the dissipationless case.
Letters (a), (b), (c) and (d) refer to the
overlaps of cavity field states for each revival depicted in fig. \ref{fig:3}. 
As expected, spontaneous revivals (b) and (c) are symmetric with respect to
the flat signal and involve only one rapid frequency, respectively given by 
$3g\sqrt{\bar{n}}$ and $2g\sqrt{\bar{n}}$. }
\end{figure}

Let us now consider the case of three atoms. 
Figure \ref{fig:5} presents
a comparison between the results of an analytic exact diagonalization 
of the Tavis-Cummings Hamiltonian and
the mesoscopic approximation. 
Correspondence with the revivals described in fig. \ref{fig:3} is indicated.
The upper and lower envelopes \eqref{eq:enveloppe:N>1:upper}
and \eqref{eq:enveloppe:N>1:lower} are depicted. Fig. \ref{fig:5} shows that although 
the effective Hamiltonian does
not fully reproduce the exact signal, it does reproduce the amplitude and the positions of
the revivals in a satisfactory way. 

As expected, eqs. \eqref{eq:enveloppe:N>1:upper}
and \eqref{eq:enveloppe:N>1:lower} effectively describe the upper and lower 
envelopes of the signal during the early collapse of Rabi oscillations. They fit
also rather well with the first complete revival ($\phi\sim 2\pi$). 
But they fail for the revival at $\phi\sim \pi$:
\eqref{eq:enveloppe:N>1:upper} 
corresponds to the upper envelope but \eqref{eq:enveloppe:N>1:lower} does not. 
This is not surprising since this revival is due to
$q=2$. In this case, formulas \eqref{eq:enveloppe:N>1:upper} and \eqref{eq:enveloppe:N>1:lower}
weigh the $q=2$ contribution with the same sign. Eq. \eqref{eq:enveloppes:N=1} with
$q_{r}=2$ would be more appropriate to describe the envelopes near this revival. 

Finally, the value 
${\mathcal A}_{0}$ of the probability between revivals obtained from the effective Hamiltonian 
differs from the one obtained from the numerical solution of the Schrödinger equation in the
case $N=3$. This is a finite $\bar{n}$ effect arising from the choice of the
effective Hamiltonian \eqref{eq:Heffective}. It can be checked that this difference
vanishes as $\bar{n}^{-1/2}$ in the classical limit $\bar{n}\rightarrow +\infty$.

\begin{figure}
\includegraphics[width=9cm]{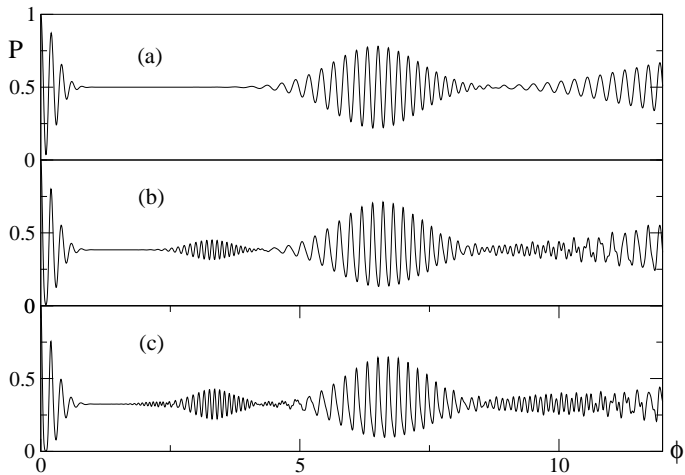}
\caption{\label{fig:2}
Spontaneous revivals of the Rabi oscillation signal for $\bar{n}=15$,
without dissipation, computed from a numerical integration of the Schrödinger
equation for (a) $N=1$ atom, (b) $N=2$ atoms, (c) $N=3$ atoms 
as a function of the dimensionless time $\phi=gt/2\sqrt{\bar{n}}$. 
The initial condition is $m=N/2$ (all atoms excited). }
\end{figure}

Comparing Rabi oscillation revival patterns at fixed $\bar{n}$ for various $N$ shows
that using two or three atoms instead of one induces an earlier spontaneous revival because
the extreme Gea-Banacloche states ($m=\pm J$) move faster than the ones associated
with $m=\pm 1/2$. But the weight of Rabi oscillations generated by high $|m|$
polarizations quickly decreases with $N$. Fig. \ref{fig:2} suggest that the
first spontaneous Rabi revival for $N=2$ and the second one for $N=3$ could be good
candidates for the observation of spontaneous Rabi revivals. Of course, 
dissipation in the cavity leads to smaller Rabi oscillations as we shall see in the next section.

\section{Dissipative dynamics}
\label{sec:dissipative}

\subsection{Stochastic wave function approach to quantum dynamics}

\subsubsection{General principle}

Within the context of cQED experiments performed with Rydberg atoms
in microwave cavities, dissipation almost exclusively originates in
cavity losses. They are extremely low since the quality factor $Q$ of the cavity is of the order or 
higher than $10^8$. Dissipation can be modeled through the coupling of the cavity mode 
to an harmonic bath with very short memory.
Within this framework, 
the dynamics of the coupled atom + cavity system can be described by a
master equation for its reduced density matrix. The master equation is valid over time scales
much larger than the memory time  $\tau_{c}$ of the bath. In the weak dissipation limit,
which is realized here, it is still valid down to 
$T\simeq 0$~K and, in the present case, takes the form:
\begin{equation}
\label{eq:master-equation}
\frac{d\rho}{dt} = -\frac{i}{\hbar}\,[H,\rho] + \gamma\,a\ldotp\rho\ldotp
a^\dagger -\frac{\gamma}{2}(a^\dagger a\ldotp \rho+\rho\ldotp a^\dagger a)
\end{equation}
where $H$ denotes the Tavis-Cummings Hamiltonian \eqref{eq:Dicke-SU2}. 
Note that switching to an interaction representation for the atoms and the cavity mode 
does not modify the form of the dissipative terms and simplifies
the Hamiltonian part.
In principle, eq. \eqref{eq:master-equation} can be solved
numerically in order to obtain the quantum dynamics. 
However, an analytical ansatz for the reduced density matrix can be
found within the mesoscopic approximation. As we shall see in the next section, 
this ansatz is conveniently derived using an alternative but equivalent approach to the dissipative
dynamics of the atoms + cavity system: the quantum jump 
approach \cite{Dalibard:1992-1}.

\medskip

The basic idea underlying this approach is to consider that the environment of the system is
continuously monitored so that any emission or absorption of quanta by the system
can be assigned a precise date. Each time such an event occurs, the system undergoes a quantum
jump. Between these jumps, its evolution is described by an effective Hamiltonian that
describes both its intrinsic dynamics and the acquisition of information arising from
the fact that no quanta has been detected. The probability rates
for the various quantum jumps are directly obtained as averages 
of $L_{i}^\dagger L_{i}$ in the state under consideration
where the $L_{i}$ denote the quantum jump
operator (here, only $L=\sqrt{\gamma}\,a$ is present at zero temperature). The 
reduced density matrix
is then recovered by averaging over the set of stochastic trajectories associated 
with a large set of quantum jumps sequences. The weight
of a given trajectory can be directly related to the dates and 
types of the various quantum jumps.
 
This method proves to be very
convenient numerically since the number of variables involved is of the order
of the dimension $d$ of the system's Hilbert space whereas it scales
as $d^2$ in the master equation approach. Note that the quantum jump approach
is the only adequate formalism for studying the behavior of a single realization of the
quantum system. 

\subsubsection{Decoherence of coherent state Schrödinger cats}
\label{sec:dissipative:coherent}

Before applying this method to our problem, it is instructive to recall how the dissipative dynamics
of an harmonic oscillator can be described in this way. In particular, we shall review how
the decoherence scenario for a superposition of two coherent states can
be recovered within this framework since it will prove to be useful in our cQED problem. 
At $T=0$~K, the system can only emit quanta. The stochastic dynamics of the quantum
state is then described as follows:
\begin{itemize}
\item During a small time interval $\tau\gg \tau_{c}$, the probability for
a quantum jump is $p_\tau=\gamma \tau \langle \psi(t)|a^\dagger a|\psi(t)\rangle $
and the state after such a jump is $|\psi(t^+)\rangle =\sqrt{\gamma\tau}\,
a\,|\psi(t)\rangle/\sqrt{p_\tau}$. 
\item Between jumps, the effective non-hermitian Hamiltonian is given by
\begin{equation}
\label{eq:nh-hamiltonian}
\hbar^{-1}H_{\mathrm{eff}}=\omega_{0}a^\dagger a-i\frac{\gamma}{2} a^\dagger a
\end{equation}
and between $t$ and $t+dt$ the state evolves according to
\begin{equation}
\label{eq:evolution:nojump}
|\psi(t+dt)\rangle= \frac{(1-i\hbar^{-1}H_{\mathrm{eff}} dt)\,|\psi(t)\rangle}{\sqrt{\Pi_{0}(t,dt)}}
\end{equation}
where the probability 
$\Pi_{0}(t,dt)$ that no quantum jump occurs between $t$ and $t+dt$ is given, in
the present situation, by $\Pi_{0}(t,dt)=1-p_{dt}$.
\end{itemize}
The evolution of a single coherent state $|\alpha\rangle $ is well known: 
apart from a well defined global phase factor, a quantum jump does not
affect it. A coherent state remains coherent during the evolution 
between quantum jumps but its amplitude decreases exponentially due to the acquisition
of information by negative measurements (no photon escapes) between the jumps: 
$\alpha(t)=\alpha\, e^{i\omega_{0}t-\gamma t/2}$. 

Let us now consider a superposition of two distinct 
coherent states with the same average number of
quanta $\bar{n}$ but with a relative phase $\theta$ in phase space:
\begin{equation}
\label{eq:coherent-state-cat}
|\psi(0^-)\rangle = \frac{1}{\sqrt{2}}\left(|\alpha \rangle +
|e^{i\theta}\alpha\rangle\right)\,.
\end{equation}
The jumps probability during time $\tau$ is then given  by
$p_\tau=\bar{n}\gamma\tau\left(
1+2\,\Re{(\langle \alpha|\alpha e^{i\theta}\rangle\,e^{i\theta})}
\right)$ which simplifies to $\bar{n}\gamma\tau$ as soon as 
$|\alpha\rangle$ and $|\alpha\,e^{i\theta}\rangle$
are well separated so that their overlap can be safely neglected. Under this assumption, the state
after a jump is given by:
\begin{equation}
\label{eq:psitplus:coherent}
|\psi(0^+)\rangle =\frac{1}{\sqrt{2}}\left(e^{i\mathrm{Arg}(\alpha)}\,|\alpha\rangle + 
e^{i(\mathrm{Arg}(\alpha)+\theta)}\,|\alpha e^{i\theta}\rangle\,.
\right)
\end{equation}
In the present unfolding of the
master equation, each quantum jump introduces a phase factor $e^{i\theta}$ in the quantum 
superposition whereas 
each component remains a coherent state with the same parameter. 
Decoherence occurs because the number of
jumps in a given time interval varies from one stochastic trajectory to the other. 
This has already been noticed in \cite{Milman:2000-1} using an
unfolding of the master equation based on a continuous time measurement
through an homodyne detection of the field leaving the cavity. The present scheme leads to 
the same final results but is more suited for our study of 
dissipation on the atoms + cavity dynamics.

Denoting by $\{t_{1},\ldots,t_{p}\}$ the successive dates of quantum jumps 
($0\leq t_{1}< \ldots < t_{p}\leq t$), the final state associated with this sequence
of jumps is given by:
\begin{equation}
\label{eq:psi:multiple-jumps}
|\psi_{\{t_{1},\ldots,t_{p}\}}(t)\rangle =
\frac{1}{\sqrt{2}}\left(|\alpha e^{-\gamma t/2}\rangle + e^{ip\theta}
|\alpha e^{-\gamma t/2} e^{i\theta}\rangle\,.
\right)
\end{equation}
Thus, the decoherence coefficient is the characteristic function for the probability 
distribution of the number of quantum jumps $N[0,t]$
between $0$ and $t$. The oscillator reduced density matrix at time $t$ is given by:
\begin{eqnarray}
\rho(t) & = & |\alpha (t)\rangle\langle\alpha (t)|
+|\alpha(t)e^{i\theta}\rangle\langle\alpha(t)e^{i\theta}| \\
& + & \mathcal{D}(t)\,|\alpha(t)e^{i\theta}\rangle\langle\alpha(t)|
+\mathcal{D}(t)^*\, |\alpha(t)\rangle\langle\alpha(t)e^{i\theta}|\nonumber
\end{eqnarray}
where $\mathcal{D}(t)$ denotes the average of $e^{iN[0,t]\theta}$ over all stochastic trajectories.
When $\gamma t\gtrsim 1$, the reduction of the components amplitude has to
be taken into account in order to get the probability distribution for
a sequence of quantum jumps dates $0\leq t_{1}\leq \ldots t_{p}\leq t$.
This step is necessary to recover 
the full solution of the master equation \eqref{eq:master-equation}.
These computations are recalled in appendix \ref{sec:appendix:diffusion}.
But for $t\ll \gamma^{-1}$, relaxation of 
energy has almost not occured and we can
assume that the average number of quanta in the two coherent
components of the superposition is still equal to $\bar{n}$. 
%The probability of having
%a quantum jump between $t'$ and $t'+\tau$ does not depend ot $t'$. 
Therefore, emission of quanta is a renewal process with a distribution of
waiting times given by $\psi(\tau)=\bar{n}\gamma\,e^{-\bar{n}\gamma \tau}$. 
Decoherence by a sequence of quantum jumps obeying a renewal process has
been recently studied in full generality \cite{Degio20}. In the present case, 
since $N[0,t]$ is distributed according to Poisson law with mean value $\bar{n}\gamma t$, the result
is given by:
\begin{equation}
\label{eq:decoherence:coherent}
\mathcal{D}(t)=\langle e^{iN[0,t]\theta}\rangle = 
\exp{\left(\bar{n}\gamma t\,(e^{i\theta}-1)\right)}\,
\end{equation}
and leads to the same predictions as the direct solution of the master equation.

\subsection{Decoherence in the mesoscopic approximation}
\label{sec:dissipative:cQED}

Let us now turn to the dissipative dynamics in the atoms + cavity problem. 
Because Rydberg atom
experiments are performed over time scales rather short compared to the dissipation time
($\gamma t\lesssim 0.1$), we shall look at the dissipative dynamics at short times
when energy dissipation can be neglected ($\gamma t\ll 1$). 
In order to make an explicit connection with
the work of Gea-Banacloche who has studied the effect of dissipation 
for one atom at arbitrary times \cite{Banacloche:1993-1}, 
the case of longer times ($\gamma t\gtrsim 1$) 
is discussed in appendix \ref{sec:dissipation}. 
Let us finally mention that the case of an atomic ensemble
has also been considered within the framework of master equation 
\cite{Saavedra:1998-1} which, in our opinion, does not clarify the
dissipative quantum dynamics of the 
cavity + atoms system as much as the quantum stochastic
trajectories method discussed below.

\subsubsection{Evolution along a single stochastic trajectory}

Inspired by the dissipationless case,
we will focus on the evolution of factorized states of the form 
\begin{equation}
\label{eq:psitildeXinit:main}
|\Psi^X_{m}\rangle = \sum_{m''=-J}^JR^{-1}_{m'',m}
|J,m''\rangle\otimes |\alpha\rangle\,,
\end{equation}
which,  within the mesoscopic approximation, remain factorized in the absence of dissipation.
Strictly speaking, as noticed by Gea-Banacloche for $N=1$ and as proved in appendix
\ref{sec:dissipation}, this is not true in the presence of dissipation. 
However, in the limit $\gamma t\ll 1$, the dissipative dynamics can still be formulated
in terms of factorized states. Let us sketch the argument that justifies
this assertion. We refer the reader to appendix \ref{sec:dissipation} for details.

Between quantum jumps,
the state of the atoms + cavity system evolves according to \eqref{eq:evolution:nojump} 
using the non-hermitian Hamiltonian \eqref{eq:nh-hamiltonian} which 
takes into account the information gained
by observing that no quanta is emitted between two jumps. 
Because the atoms and the cavity mode are coupled, the whole atoms + cavity state
should be affected by this information gain. But, in the present strong coupling situation
($g\gg \gamma$), we expect the atoms to be mainly driven by the cavity and not by this indirect
information gain. 
Next, dissipation induces an exponential decay of the average photon number
$\bar{n}(t)=\bar{n}\,e^{-\gamma t}$ while keeping the photon number distribution 
Poissonian. We shall thus neglect the decay of $\bar{n}(t)$ for $\gamma t\ll 1$.
The resulting evolution for the atoms + cavity case is the same 
as in the dissipationless case:
\begin{equation}
\label{eq:psitilde:factorized:main}
|\Psi^X_{m}(t)\rangle\simeq e^{-igmt\sqrt{\bar{n}}}\,
|D_{m}(t)\rangle\otimes|\psi_{m}(t)\rangle\;.
\end{equation}
Let us now discuss the effect of a quantum jump on this state.
Contrarily to coherent states, each Gea-Banacloche state
$|\psi_{m}(t)\rangle $ does not remain invariant under the action of a 
quantum jump operator since:
\begin{eqnarray}
\label{eq:afterjump:GBstate-1}
 a\,|\psi_{m}(t)\rangle & = & e^{-\bar{n}/2}\sum_{k=0}^\infty
 \alpha\,\frac{\alpha^k}{k!}\,e^{-igmt\sqrt{k+1}}\,|k\rangle\\
 \label{eq:afterjump:GBstate-2}
 & = & e^{-\bar{n}/2}\sum_{k=0}^\infty
 \alpha \, e^{imgt(\sqrt{k}-\sqrt{k+1})}
 \,\frac{\alpha^k}{k!}\,e^{-igmt\sqrt{k}}\,|k\rangle
\nonumber
\end{eqnarray}
The phase factor $\exp{[imgt(\sqrt{k+1}-\sqrt{k})]}$ {\sl a priori} depends upon $k$. 
But expanding $\sqrt{k+1}-\sqrt{k}$ in powers of $(k-\bar{n})/\sqrt{\bar{n}}$
shows that, at first order, 
$\exp{(imgt(\sqrt{k+1}-\sqrt{k}))}$ is indeed independent of $k$. Using
this approximation, the action of the annihilation operator reduces to the multiplication by
a phase:
\begin{equation} 
\label{eq:afterjump:GBstate}
a\, |\psi_{m}(t)\rangle \simeq \alpha\, e^{-imgt/2\sqrt{\bar{n}}}
\, |\psi_{m}(t)\rangle\,.
\end{equation}
It can be shown that an expansion to the next order differs from this expression by
$O((mgt/\bar{n})^2)$. Thus, \eqref{eq:afterjump:GBstate} can be considered
as a valid approximation in the domain $t\ll \sqrt{\bar{n}}/g$ which is in
the domain of validity of the mesoscopic approximation ($t\ll \bar{n}/g$).

\medskip

Let us now consider the evolution of state \eqref{eq:psitildeXinit:main} along a single stochastic
trajectory. Just before its first quantum jump, provided it happens at time $t_{1}$ such
that $\bar{n}(t_{1})\gg N$, we still have a factorized state of the 
form \eqref{eq:psitilde:factorized:main}. The 
effect of a quantum jump occuring a time $t_{1}$ is to extract a phase $e^{i\theta_{m}(t_{1})}$
where $\theta_{m}(t_{1})$ is the argument of the amplitude of the 
quasi-coherent state $|\psi_{m}(t_{1})\rangle$. 
Iterating this argument shows that,
in a stochastic trajectory with quantum jumps occuring at times
$0\leq t_{1}<\ldots <t_{p}\leq t$, $|\Psi_{m}^X\rangle$ remains factorized
but gets an extra phase $\theta_{m}(t_1,\ldots,t_{p})$
associated with the quantum jumps:
\begin{equation}
\label{eq:psitilde:factorized:stochastic}
|\Psi^X_{m}\{t_{1},\ldots,t_{p}\};\,t\rangle=
e^{i\theta_{m}(t_{1},\ldots,t_{p})}\,e^{-igmt\sqrt{\bar{n}}}
|D_{m}(t)\rangle \otimes |\psi_{m}(t)\rangle
\end{equation}
and $\theta_{m}(t_1,\ldots,t_{p})=\sum_{j}\theta_{m}(t_{j})$.
Exactly as for the case of a superposition of coherent states of an harmonic oscillator analyzed above,
the accumulation of random relative phases in front of 
the Gea-Banacloche states 
leads to the decoherence of the Schrödinger cat state created by the strong atoms + cavity
coupling and to the disappearance of Rabi oscillations. 

\medskip

Before discussing the average over all stochastic trajectories, it is worth mentioning 
that the present discussion remains valid even in the presence of the echo pulses
introduced in section \ref{sec:Banacloche}. A $\pi$-pulse 
instantaneously reverses the dynamics
of the atoms + cavity system. 
After a single $\pi$-pulse at time $t_{\pi}$, the Gea-Banacloche states invert their
motion and start refocusing. The deterministic evolution of the atoms + cavity system
is then described by a time-reversed evolution of the dissipationless motion. Therefore, the
effect of any subsequent quantum jump is still to extract a phase corresponding to the
position of the quasi-coherent Gea-Banacloche state in the Fresnel plane at the jump time. 

\subsubsection{Average over stochastic trajectories}
\label{sec:Monte-Carlo:average}

To deal with all these situations at once, let us 
denote by $\theta_{m}(t)$ the time-dependent phase of $|\psi_{m}(t)\rangle$,
not assuming any particular form. 
The decoherence coefficient for the two states 
$|\Psi^X_{m_{\pm}}(t)\rangle$ considered here is thus given 
by the average over sequences of quantum jumps:
\begin{equation}
\label{eq:influence-definition}
\mathcal{F}[\theta_{m_{+}},\theta_{m_{-}}]=\Big\langle
e^{i\sum_{l}(\Delta\theta)(t_{l})}
\Big\rangle
\end{equation}
where $(\Delta\theta)=(\theta_{m_{+}}-\theta_{m_{-}})(t)$ and
the $t_{l}$ are the dates of the successive quantum jumps occuring between $0$ and $t$.
This coefficient now depends in a functional way on the two trajectories
$t\mapsto \theta_{m_\pm}(t)$ and should be called a decoherence
functional in reference to the work of Feynman and Vernon \cite{Feynman:1963-1}. 
Its definition \eqref{eq:influence-definition} generalizes \eqref{eq:decoherence:coherent}
to the case of a time dependent $\Delta\theta$. 
Since we assumed that $\gamma t\ll 1$, the statistics of waiting times between
quantum jumps
is independent of the positions of the Gea-Banacloche states. Exactly as in section 
\ref{sec:dissipative:coherent}, it
is given by $\psi(\tau)=\bar{n}\gamma e^{-\bar{n}\gamma \tau}$. 
Within this approximation,
\eqref{eq:influence-definition} can be computed explicitely even for a time dependant
$\Delta\theta$. 
An elegant way to get the result consists into rewriting
the sum over the number of quantum jumps in a completely different way which does not
singularizes any specific time:
\begin{equation}
\label{eq:magic}
\mathcal{F}[\theta_{m_{+}},\theta_{m_-}]=\Big\langle
\prod_{0\leq t'\leq t}\left(1+n(t')(e^{i(\Delta\theta)(t')}-1)\right)
\Big\rangle
\end{equation}
where $n(t')=0$ if no event occurs at time $t'$ and $n(t')=1$ when a quantum jump occurs at
time $t'$ and $(\Delta\theta)=\theta_{m_{+}}-\theta_{m_{-}}$. Note that in the
above expression, $t'$ is not the time of a quantum jump.
The formal infinite product in the r.h.s of \eqref{eq:magic} 
can then be expanded leading to an expansion 
involving multi-time correlators $\langle n(t'_{1})\,\ldots n(t'_{r})\rangle$
where $0\leq t'_{1}\leq \ldots \leq t'_{r}\leq t$ (here, 
$r$ is not the number of quantum jumps). Because 
$n(\tau)=0$ or $1$:
\begin{equation}
\langle n(t'_{1})\,\ldots n(t'_{r})\rangle = 
\langle n(t'_{1})\rangle \times \prod_{l=1}^{r-1}
\mathrm{P}\left(t'_{l+1}|t'_{l}\right)\,.
\end{equation}
where $\mathrm{P}\left(t'_{l+1}|t'_{l}\right)=
\mathrm{Prob}\left(n(t'_{l+1})=1|n(t'_{l})=1\right)$.
At short times, 
quantum jumps provide a renewal process and therefore
the conditional probabilities $\mathrm{P}\left(t'_{l+1}|t'_{l}\right)$ are 
directly related to
the average density of jumps $S(t)$:
$\mathrm{P}\left(t'_{l+1}|t'_{l}\right)=S(t'_{l+1}-t'_{l})$. Moreover, 
$\langle n(t'_{1})\rangle = S(t'_{1})$. Here the density of
events is a constant: $S(t)=\gamma \bar{n}$.
The final result for the decoherence functional is thus:
\begin{equation}
\label{eq:decoherence:short-time}
\mathcal{F}[\theta_{m_+},\theta_{m_-}] =
 \exp{\left(\gamma\,\bar{n}\int_{0}^t
(e^{i(\Delta\theta)(\tau)}-1)\,d\tau \right)}\,.
\end{equation}
Finally, expression \eqref{eq:decoherence:short-time} can be interpreted as 
resulting from the accumulation of decoherence 
coefficients over infinitesimal periods of time. In the mesoscopic regime, because
of the strong coupling
regime, the evolution of the atoms + cavity system is a forced evolution 
of the cavity state driven by the atomic polarizations
$|D_{m}(t)\rangle$, leading to the motion of the Gea-Banacloche states
$|\psi_{m}(t)\rangle$.
For each Gea-Banacloche state, this forced motion between $\tau$ and $\tau+d\tau$
leaves an imprint in the environment of the cavity. The overlap between imprints
left by two distinct Gea-Banacloche states is precisely
the decoherence coefficient.
During infinitesimal time $d\tau$, the imprint left
in the environment by each
Gea-Banacloche state under consideration is the same as the one left
by coherent states with the 
same average number of quanta but a time-dependent phase separation $(\Delta\theta)(\tau)$.
Therefore, the corresponding decoherence coefficient is given by:
\begin{equation}
\label{eq:infdecoh}
D(\tau,\tau+d\tau)\simeq 
\exp{\left(\gamma \bar{n}\,(e^{i(\Delta\theta)(\tau)}-1)\,d\tau\right)}\,.
\end{equation}
Since the environment is Markovian, 
the infinitesimal decoherence coefficients \eqref{eq:infdecoh} 
associated with
different time windows $[t,t+\tau]$ ($\tau\gg \tau_{c}$)
accumulate through time evolution, leading to 
\eqref{eq:decoherence:short-time}.

\subsection{Rabi oscillations in the presence of dissipation}
\label{sec:analytical}

\subsubsection{Spontaneous revivals}
\label{sec:analytical:free}

It is now straightforward to compute the Rabi oscillation signals by introducing the
decoherence coefficients for all pairs of Gea-Banacloche states that can appear in the reduced atomic
density operator. Because we are dealing with experimental situations such that
the duration of experiment is small compared to the cavity dissipation time, 
we shall assume that the average number of quanta remains equal to $\bar{n}$ 
in this paragraph and the following ones.

\medskip 

The result for the Rabi oscillation signal 
is ($q=m_{+}-m_{-}$):
\begin{equation}
\label{eq:Rabi:free:general}
P_{m}(t)=\sum_{m_{+},m_{-}}e^{-igqt\sqrt{\bar{n}}/2}
\mathcal{P}_{m_{+},m_{-}}(t)\,
\mathcal{R}_{m_{+},m_{-}}(t)\,\mathcal{F}_{m_{+},m_{-}}(t)
\end{equation}
where $\mathcal{P}_{m_{+},m_{-}}(t)$ and $\mathcal{R}_{m_{+},m_{-}}(t)$ 
are respectively given by equation \eqref{eq:defP} and \eqref{eq:defR}.  
Decoherence is contained in 
$\mathcal{F}_{m_{+},m_{-}}(t)=e^{-d_q(t)+i\Theta_q(t)}$
which can be evaluated using \eqref{eq:decoherence:short-time}, thus leading to
 ($\phi=gt/2\sqrt{\bar{n}}$):
\begin{eqnarray}
d_{q}(t) & = & 
\frac{2\gamma \bar{n}^{3/2}\,\phi}{g}
\,\left(
1-\frac{\sin{(q\phi)}}{q\phi}
\right) \label{eq:free:decay}\\
\Theta_{q}(t) & = & \frac{4\gamma\,\bar{n}^{3/2}}{g\,q}\,
\sin^2{\left(\frac{q\phi}{2}\right)}\,.
\label{eq:free:phase}
\end{eqnarray}
Note that in these results, $\bar{n}^{3/2}\gamma/g$ 
is the dimensionless parameter that characterizes the
strength of decoherence. 

An ansatz for the upper and lower envelopes
of the Rabi oscillation signal in the presence of dissipation can then be obtained along
the lines of section \ref{sec:effectiveH:enveloppes}:
\begin{eqnarray}
\label{eq:dissipative:enveloppe:N>1:upper}
P_{+}(t) & = & \mathcal{A}_{0}+\sum_{q\neq 0}
|\mathcal{R}_{q}(t)\,\mathcal{A}_{q}|\,e^{-d_{q}(t)}\\
\label{eq:dissipative:enveloppe:N>1:lower}
P_{-}(t) & = & \mathcal{A}_{0}+\sum_{q\neq 0}(-1)^q
|\mathcal{R}_{q}(t)\,\mathcal{A}_{q}|\,e^{-d_{q}(t)}\,.
\end{eqnarray}

\subsubsection{Induced revivals}
\label{sec:analytical:induced}

Rabi oscillation signals in an echo experiment can also be computed within the
mesoscopic approximation. The percussional echo pulse is applied to the atoms at time
$t_{\pi}$. It corresponds to the operator 
$U_{\pi}=\otimes_{j=1}^N\sigma_{j}^z$.  Using this operator, equation \eqref{eq:echo:evolution:result}
can be derived for the case of $N$ atoms with $H_{TC}$ in place of the Jaynes-Cummings Hamiltonian.
Thus, exactly as for $N=1$, the
evolution of the atoms + cavity system is reversed after time $t_{\pi}$. Within the mesoscopic approximation,
this means that atomic polarization as well as Gea-Banacloche states move backward towards their
initial positions. The time dependence of the
associated phases $\theta_{m}(\tau)$ associated with Gea-Banacloche states
is given by:
\begin{equation}
\begin{cases}
\theta_{m}(\tau)=mg\tau/2\sqrt{\bar{n}}\quad\mathrm{for}\ 0\leq \tau\leq t_{\pi}\\
\theta_{m}(\tau)=mg(2t_{\pi}-\tau)/2\sqrt{\bar{n}}\quad\mathrm{for}\ t_{\pi}\leq \tau
\end{cases}
\end{equation}
Eq. \eqref{eq:decoherence:short-time} 
leads to the following decoherence coefficient ($t\geq t_{\pi}$)
which we write as $\mathcal{F}_{m_{+},m_{-}}^{(\mathrm{echo})}(t_{\pi},t) =
e^{-d_{q}(t_{\pi},t)+i\Theta_{q}(t_{\pi},t)}$ where ($q=m_{+}-m_{-}$):
\begin{equation}
d_q(t_{\pi},t) = \frac{2\gamma\,\bar{n}^{3/2}}{g}\,\left(\phi-
\frac{
2\sin{\left(q\,\phi_{\pi}\right)}
 -  \sin{\left(q\,(2\phi_{\pi}-\phi)\right)}}{q}\right)
\label{eq:echo:decay}
\end{equation}
where $\phi_{\pi}=gt_{\pi}/2\sqrt{\bar{n}}$ and
\begin{equation}
\Theta_q(t_{\pi},t) = \frac{4\gamma\,\bar{n}^{3/2}}{g\,q}\,\left(
2\sin^2{\left(q\phi_{\pi}/2\right)}-
\sin^2{\left(q(2\phi_{\pi}-\phi)/2\right)}
\right)\,. 
\label{eq:echo:phase}
\end{equation}
Because of the perfect time reversal for
the atoms + cavity system, the overlap factor $\mathcal{R}_{m_{+},m_{-}}$ 
in the echo experiment can be expressed in terms of
the overlap factor under free evolution for $t\geq t_{\pi}$:
$\mathcal{R}^{(\mathrm{echo})}(t_{\pi},t)=\mathcal{R}(2t_{\pi}-t)$.

\subsubsection{Extension to finite temperature}
\label{sec:analytical:temperature}

It is known that increasing the temperature lowers the decoherence time. 
For the harmonic oscillator, initially in a
coherent state, the exact solution to the quantum master equation 
\eqref{eq:master-equation} is well known \cite{Kim:1992-1}. At time $t$, the state
is no longer pure but appears to be a thermal state with average number
of quanta $\bar{n}(t)=\bar{n}_{T}(1-e^{-\gamma t})$ translated in phase space
by $\alpha e^{-\gamma t/2}$ ($\bar{n}_{T}$ denoting the average number of quanta at equilibrium at
temperature $T$). Nevertheless for times much shorter than dissipation, thermalization
can be neglected: the state of the oscillator can still be considered as coherent. The analysis of the
solution of the master equation at finite temperatures shows that decoherence
at short times is still exponential. The effect of temperature is to enhance the
damping rate by a factor $2n_{\mathrm{th}}+1=\coth{(\beta\hbar\omega_{0}/2)}$.
This suggests that the imprint of the superposition of two coherent
states in the environment during an infinitesimal time interval $d\tau $ at finite temperature
is obtained by substituting $\gamma \rightarrow \gamma \coth{(\beta\hbar\omega_{0}/2)}$
in eq. \eqref{eq:infdecoh}. Following the previous line of reasoning (end of section 
\ref{sec:Monte-Carlo:average}), 
the decoherence coefficient at time $t$ for a superposition of two coherent
states is again obtained by summing the decoherence coefficients 
associated with infinitesimal time intervals between $\tau=0$ and $\tau=t$. 

\medskip

This result can be used to derive the evolution of the atoms + cavity density matrix
at short times and moderate finite temperature. 
As long as we can neglect the thermalization, the only effect of dissipation is to damp 
the coherences
between states $|D_{m_{\pm}}(t)\rangle\otimes |\psi_{m_{\pm}}(t)\rangle$ for 
$m_{+}\neq m_{-}$. As in the zero temperature case,  
each of the state $|D_{m_{\pm}}(t)\otimes |\psi_{m_{\pm}}(t)\rangle$ is expected to
evolve according to the atoms + cavity interaction and the 
echo pulse applied to the system (if any). As explained in the previous paragraph,
the decoherence coefficient
$\mathcal{F}_{m_{+},m_{-}}(t)$ to be used in eq. \eqref{eq:Rabi:free:general} is obtained by 
replacing $\gamma$ by $\gamma\,\coth{(\beta\hbar\omega_{0}/2)}$ in equations \eqref{eq:free:decay},
\eqref{eq:free:phase} for the free evolution and
\eqref{eq:echo:decay}, \eqref{eq:echo:phase} for
the echo experiment. Note that this ansatz is expected to be valid only for
low temperatures and at short times such that $\gamma t\,
\coth{(\beta\hbar\omega_{0}/2)}\,\ll 1$. 

\section{Discussion of the results}
\label{sec:results}

\subsection{Method and parameters}
\label{sec:results:parameters}

We have considered 
the Rabi oscillation signal in the presence of dissipation for $N=1$ to $N=3$ atoms, 
values that can be realistically reached in state-of-the-art cavity 
QED experiments. Photon
numbers $\bar{n}=10$ and $15$ have been considered.
All our computations have been performed for values of $g/\gamma$ corresponding
to the present ENS experiment \cite{Meunier:2005-1}.  The best 
published cavity damping time is 1~ms (quality factor $Q=3.2\times 10^8$). 
Preliminary tests of an 
improved experimental setup have shown damping rates of 14~ms ($Q=4.5\times10^9$)
and even 115~ms ($Q=3.7\times10^{10}$) and these results are to be submitted in
the near future. 
%For this last
%value, cavity damping is no longer the most important source of dissipation 
%since spontaneous emission which we have neglected, takes place in 30~ms.
Thus, the values of 14~ms ($g/\gamma\simeq 4310$),
5~ms ($g/\gamma\simeq 1540$ and $Q=1.6\times 10^9$) and 1~ms ($g/\gamma\simeq 308$) 
for the damping rate of the cavity have been considered in our simulations.
We focus on the case of a zero-temperature bath which can be realistically reached as shown in \cite{Brune:1996-1}.
The effect of finite temperature will be briefly discussed in 
section \ref{sec:results:temperature}.

\medskip

Results of the analytical approach described in section \ref{sec:analytical} 
have been compared to a quantum Monte-Carlo simulation 
of the atoms + cavity system evolution
in the spirit of \cite{Dalibard:1992-1}. For these simulations, the Adams-Bashford
scheme of order four has been used to compute the evolution of the wave function between quantum
jumps. 

\medskip

We first present our results relative to the free evolution of the atoms + cavity (spontaneous Rabi 
oscillations revivals) 
in section \ref{sec:results:free} and for the echo experiments in section
\ref{sec:results:echo}. Consequences of these results for cQED and circuit-QED experiments
are then discussed in section \ref{sec:results:discussion}.

\subsection{Free evolution}
\label{sec:results:free}

\begin{figure}
\includegraphics[width=8cm]{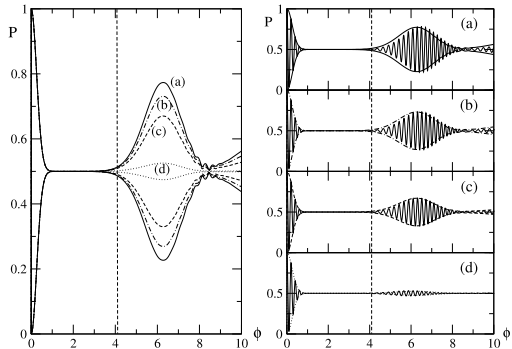}
\caption{\label{fig:6N1}
Influence of dissipation on the spontaneous revivals of the Rabi oscillation signal
$P=P_{m=J}$ (as a function of  $\phi=gt/2\sqrt{\bar{n}}$) for one atom and 
$m=1/2$ (atom excited) with $\bar{n}=15$ photons initially. 
The graph on the left depicts the 
analytical envelopes $P_{\pm}(t)$ for
(a) no dissipation, (b) $\gamma^{-1}\simeq 14$~ms,
(c) $\gamma^{-1}\simeq 5$~ms and (d) $\gamma^{-1}\simeq 1$~ms. The right  
part of the figure presents the associated Rabi oscillation signals obtained from 
quantum Monte-Carlo simulations (plain lines) as well as the associated analytical 
enveloppes.
The vertical dashed line corresponds to the largest 
reachable $\phi$ (atoms at $100\ m\,s^{-1}$).}
\end{figure}

\begin{figure}
\includegraphics[width=8cm]{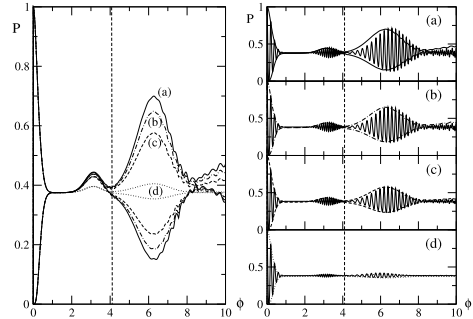}
\caption{\label{fig:6N2}
Influence of dissipation on the spontaneous revivals of the Rabi oscillation signal $P=P_{m=J}$ 
(as a function of  $\phi=gt/2\sqrt{\bar{n}}$) for $N=2$ atoms. The initial condition is
$m=1$ (all atoms excited)
with $\bar{n}=15$ photons.  The graph on the left depicts the 
analytical envelopes $P_{\pm}(t)$ for
(a) no dissipation, (b) $\gamma^{-1}\simeq 14$~ms,
(c) $\gamma^{-1}\simeq 5$~ms and (d) $\gamma^{-1}\simeq 1$~ms. The right  
part of the figure presents the associated Rabi oscillation signals obtained from 
quantum Monte-Carlo simulations (plain lines) as well as the associated analytical 
enveloppes.
%The vertical dashed line corresponds to the largest 
%reachable $\phi$ (atoms at $100\ m\,s^{-1}$).
}
\end{figure}

\begin{figure}
\includegraphics[width=8cm]{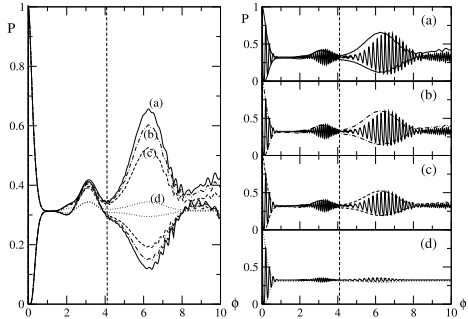}
\caption{\label{fig:6N3}
Influence of dissipation on the spontaneous revivals of the Rabi oscillation signal
$P=P_{m=J}$ 
(as a function of  $\phi=gt/2\sqrt{\bar{n}}$) for $N=3$ atoms. The
initial condition is $m=3/2$ (all atoms excited)
with $\bar{n}=15$ photons.  The graph on the left depicts the 
analytical envelopes $P_{\pm}(t)$ for
(a) no dissipation, (b) $\gamma^{-1}\simeq 14$~ms,
(c) $\gamma^{-1}\simeq 5$~ms and (d) $\gamma^{-1}\simeq 1$~ms. The right  
part of the figure presents the associated Rabi oscillation signals obtained from 
quantum Monte-Carlo simulations (plain lines) as well as the associated analytical 
enveloppes.
} 
\end{figure}

It is interesting to assess the possibility of observing spontaneous Rabi oscillation revivals, 
since, for $N>1$, such a revival might be observable at shorter
times than in the $N=1$ case. Figure \ref{fig:6N1} presents a comparison between
the Rabi oscillation signals resulting from the interaction with a mesoscopic
coherent state containing 15 photons in average in the dissipationless case and for dissipation times
equals to $1$~ms, $5$~ms and $14$~ms ($T=0$~K).
Figures \ref{fig:6N2} and \ref{fig:6N3} present the same comparison for the cases of $N=2$ and
$N=3$ atoms respectively. Note that our analytical model (eqs. \eqref{eq:dissipative:enveloppe:N>1:upper} 
and \eqref{eq:dissipative:enveloppe:N>1:lower})
predicts the upper and lower envelopes
of the Rabi oscillation signal with a rather good precision in the presence of dissipation.
Note that in the $N=1$ case, this simulation shows that it is not possible to observe 
spontaneous Rabi oscillation revivals with the present cavity ($\gamma \lesssim 1$~ms).
The same conclusion is valid for $N=2$ and $N=3$: even the partial revivals that occur
before the main one $\phi\sim 2\pi$ should not be observable.
For $\gamma^{-1} =1$~ms, 
decoherence transforms the entangled atom + cavity state into a statistical
mixture before any pair of Gea-Banacloche components of the field overlap.
As figures \ref{fig:6N1} to \ref{fig:6N3} show, improving the quality factor of the cavity 
could enable a direct observation of the spontaneous revivals in the presence of $\bar{n}=15$ photons. 

\medskip

However, the experimental
apparatus sets a tight limitation on the interaction time of the atoms with the cavity field.
The number of atoms flying through the apparatus at velocities lower than $100\ m\,s^{-1}$ is too small 
in a thermal atomic beam to be used in practice. Atomic spontaneous emission is another 
limitation for very slow atoms. We stick here to the available apparatus and set 
an upper limit on $\phi$ which could roughly be estimated
as $\phi_{\mathrm{m}}\simeq 2\pi\,(2.5/\sqrt{\bar{n}})$. This upper limit for
$\phi/2\pi$ ranges from $0.65$ ($\bar{n}=15$) 
to $0.76$ ($\bar{n}=10$) which excludes the observation of the complete 
revival obtained when all Gea-Banacloche states overlap again. 
Nevertheless, as can be seen from graphs (b) and (c) 
on figures \ref{fig:6N2} and \ref{fig:6N3}, observing a spontaneous
partial revival may be possible for $N=2,3$. In the case of two atoms, the signal would correspond to
the overlap of states $|\psi_{1}\rangle$ and $|\psi_{-1}\rangle$. In the case of three atoms,
the signal would be dominated by two overlaps corresponding to $|\psi_{1/2}\rangle$ and
$|\psi_{-3/2}\rangle $ on one side and  $|\psi_{-1/2}\rangle$ and
$|\psi_{3/2}\rangle $ on the other side. Note that the partial revival associated with the overlap 
between $|\psi_{3/2}\rangle$ and $|\psi_{-3/2}\rangle$ is not within reach. Observing a 
spontaneous revival requires
$\phi\simeq \pi$ to remain within reach for the slowest atoms. This puts an upper limit 
on the average photon number close to 25 photons.

\begin{figure}
\includegraphics[width=8cm]{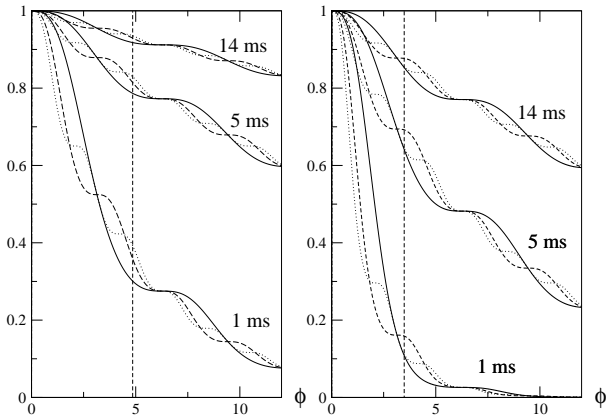}
\caption{\label{fig:decoherence-coeffs}
Modulus of the decoherence coefficient 
$\mathcal{F}_{q}(t)$ for $q=1$ (plain lines), $q=2$ (dashed lines)
and $q=3$ (dotted lines) as a function of $\phi$ 
for dissipation times $14$~ms, $5$~ms and
$1$~ms in the case of 
$\bar{n}=10$ photons (left graph) and $\bar{n}=20$ photons (right graph).
The vertical dashed line corresponds to the largest 
reachable $\phi$ (atoms at $100\ m\,s^{-1}$).
}
\end{figure}

Finally, figure \ref{fig:decoherence-coeffs} shows the decay of the modulus of the three
decoherence coefficients $\mathcal{F}_{q}(t)$ as a function of time for $\bar{n}=10$ and $\bar{n}=20$
for dissipation times equal to 1, 5 and 14~ms. It clearly shows that, even if the time of flight 
problem could be circumvented,
increasing the number of photons puts strong constraints on the dissipation time
of the cavity. This suggests that working with 10 to 15 photons in average is a good
compromise for observing spontaneous partial revivals.

\subsection{Echo experiment}
\label{sec:results:echo}

\begin{figure}
\includegraphics[width=8cm]{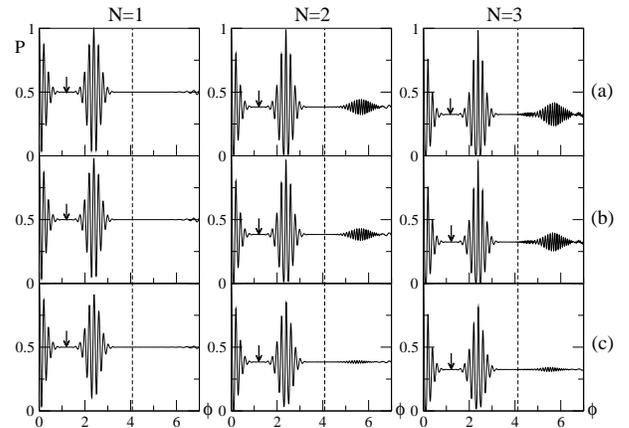}
\caption{\label{fig:echos}
Echo signal 
(function of  $\phi=gt/2\sqrt{\bar{n}}$) simulated for $\bar{n}=15$ at $t_{\pi}=30\ \mu\mathrm{s}$
and $g/2\pi=49$~kHz. All graphs on a row correspond to the same dissipation time:
(a) $\gamma^{-1}=14$~ms, 
(b) $\gamma^{-1}=5$~ms and
(c) $N=3$ and $\gamma^{-1}=1$~ms. All graphs in a column correspond to the
same numbre of atoms.
The arrow shows the 
time of the echo pulse. The vertical dashed line corresponds to the largest reachable $\phi$ value
(slowest atoms at 100 $m\,s^{-1}$).
The initial condition is $m=3/2$ (all atoms excited).} 
\end{figure}

In this paragraph, results for the simulation of an echo experiment corresponding
to an echo pulse at $t_{\pi}=30\ \mu\mathrm{s}$ are presented. Figure
\ref{fig:echos} shows the simulated echo signals for the cQED experiment at LKB
obtained with 15 photons initially and an echo pulse at 30~$\mu$s
for $N=1$, $N=2$ and $N=3$ atoms and two different values of dissipation: 14~ms, 5~ms and
1~ms. Note that in the case of three atoms,  a revival occurs
at 150~$\mu$s. It corresponds to a delayed revival of type (c) on figure \ref{fig:3}.
But although its amplitude makes it visible with a  14~ms dissipation time, the
time of flight limitation will prevent it from being observed in the experiment.
The same conclusion holds for the delayed revival predicted also in the 
$N=2$ atoms case.

\subsection{Finite temperature}
\label{sec:results:temperature}

\subsubsection{Thermalization procedure}

Experiments with Rydberg atoms in high-quality microwave cavities are
performed at low temperatures ($T\simeq 0.8-1.4$~K). In order to wash out
photons resulting from thermal leaks, an erasing procedure using auxiliary atomic samples
is applied \cite{Raimond:2001-1}.
Once the erasing procedure has been completed, a coherent field is injected inside
the cavity. Because of the imperfections of the procedure and because of 
the necessary delay $\tau_{1}\simeq 200\ \mu$s between the erasing sample 
and the coherent field
injection, this creates a translated thermal state, partially thermalized. This state would then
evolve during time $\tau_{2}\simeq 50\ \mu$s before the experimental 
atomic sample enters the cavity. 

In order to model this preparation, we have 
performed a quantum Monte-Carlo simulation
involving a finite temperature environment and an initial thermalization 
period of duration $t_{p}$. 
At the beginning of this preparation period,
a coherent state is injected in the dissipative cavity 
and evolves decoupled from the atoms during time $t_{p}$. Then
the coupling to the atoms is turned on to model the experiment.
A rather pessimistic estimate of the average number of thermal photons per mode of the reservoir has been
used for this simulation ($n_{\mathrm{th}}
\simeq 0.4$ corresponding to $T\simeq 2$~K at 51~GHz). It has been estimated that thermal
fluctuations left by the imperfect erasure procedure correspond to at most 
$n_{0}\simeq 0.15$ photons per mode on the average. Using $n_{0}=n_{\mathrm{th}}\,
(1-e^{-\gamma t_{p}})$, this sets $\gamma t_{p}\simeq 0.47$ used
in our numerical thermalization protocol. 
The injected coherent state before thermalization
has an amplitude $\alpha_{0}=\sqrt{\bar{n}}\,e^{\gamma t_{p}/2}$ in 
order to take into account the exponential decay during the preparation phase 
($\bar{n}$ denotes
the average photon number when the experimental atomic sample in injected 
inside the cavity). 

\subsubsection{Finite temperature results}
\label{sec:results:discussion}

\begin{figure}
\includegraphics[width=9cm]{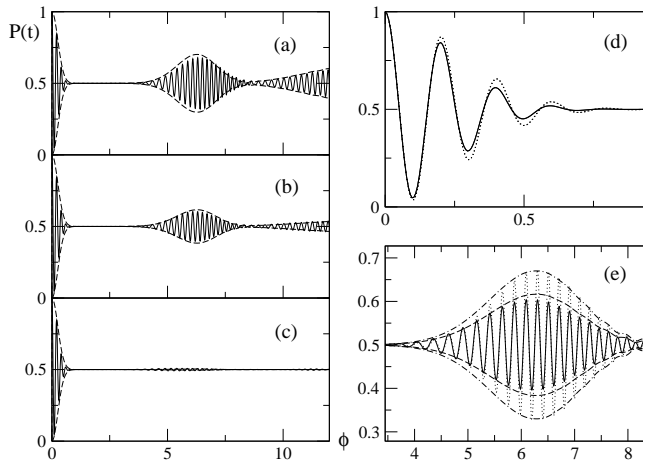}
\caption{\label{fig:thermal:1}
Rabi oscillation signals obtained from quantum Monte-Carlo simulations for
one atom using the thermalization protocol producing a thermal state with
$0.15$ photons displaced in Fresnel plane by an amplitude corresponding to $\bar{n}=15$ photons,
for (a) $\gamma^{-1}=14$~ms, (b) 5~ms and (c) 1~ms.
Graph (d): comparison between zero temperature signals (dashed line) and
finite temperature signals (full line) at short times for $\gamma^{-1}=5$~ms. 
Graph (e): comparison between
zero temperature signals (full line) and finite temperature signals (dotted line)
for $\gamma^{-1}=5$~ms. Dashed lines represent the upper and lower envelopes
$P_{\pm}$ obtained by taking temperature into account by rescaling
$\gamma \mapsto (1+2n_{\mathrm{th}})\,\gamma$. 
}
\end{figure}

Figure \ref{fig:thermal:1} presents the results of these simulations for $N=1$ atom.
The main curve presents the Rabi oscillation signals obtained from
a quantum Monte-Carlo simulation implementing 
the thermalization procedure described above
for $\gamma^{-1}=1$, 5 and 14~ms. In order to compare it with the analytical model
we have used the fact that, in the present case, at times short compared to
the dissipation time, the main effect of finite temperature is to speed
up decoherence. This is taken into account by replacing $\gamma$
by $\gamma(1+2n_{\mathrm{th}})$ in the decoherence functional 
(eqs. \eqref{eq:free:decay} and \eqref{eq:free:phase}).
This ansatz is used to compute the upper and lower envelopes that
appear on fig. \ref{fig:thermal:1}.

Note that some features are not reproduced by our analytical 
ansatz since the overlap factor we use does not take into account thermalization of
the quasi-coherent state $|\psi_{m}(t)\rangle$. 
As shown in graph (d), the initial collapse of Rabi oscillations occurs
earlier than at zero temperature. On the other hand,
the envelope of the spontaneous revival is well described by our
model (see graph (e)). This shows that our analytical approach is quite efficient 
in predicting the contrast of
spontaneous revivals of the Rabi oscillation signals even at finite temperature.

\medskip

\begin{figure}
\includegraphics[width=9cm]{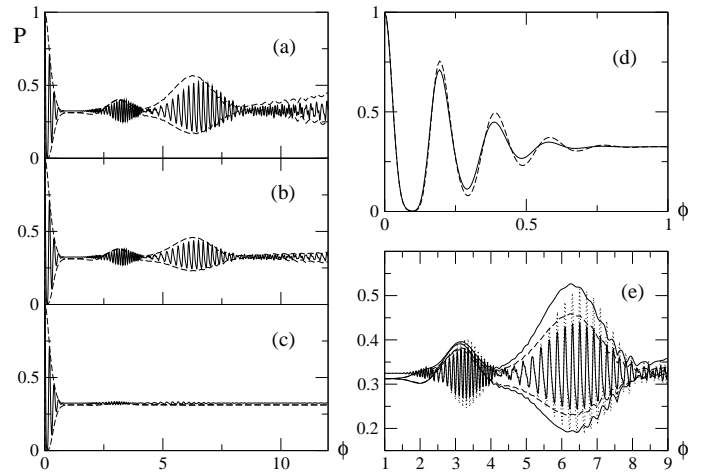}
\caption{\label{fig:thermal:3}
Rabi oscillation signals obtained from quantum Monte-Carlo simulations 
for $N=3$ atoms using the thermalization protocol producing a thermal state with
$0.15$ thermal photon number displaced in Fresnel plane with $\bar{n}=15$
for (a) $\gamma^{-1}=14$~ms, (b) 5~ms and (c) 1~ms.
Graph (d): comparison between zero temperature signals (dashed line) and
finite temperature signals (full line) at short times for $\gamma^{-1}=5$~ms. Graph (e): comparison between
zero temperature signals (full line) and finite temperature signals (dotted line)
for $\gamma^{-1}=5$~ms. Dashed lines represent the upper and lower envelopes
$P_{\pm}$ obtained by taking temperature into account by rescaling
$\gamma \mapsto (1+2n_{\mathrm{th}})\,\gamma$.
}
\end{figure}

Results for the case of three atoms are presented on figure \ref{fig:thermal:3}. The same effects
as for $N=1$ can be observed here. Note that thermal fluctuations do reduce the
contrast of spontaneous revivals albeit not enough to make them unobservable
for $\gamma^{-1}=5$ and $14$~ms. As shown in graph (e), thermal
effects reduce the contrast of the main secondary revival from 15 to 11~\% for 5~ms  dissipation time.
As mentioned above, improving the dissipation time reduces the impact of 
thermalization.

\subsection{Discussion of the results}

\subsubsection{Perspectives for the ENS experiment}

Within the context of cQED experiments performed at ENS, our results
suggests that spontaneous revivals of Rabi oscillations could be observed in an improved
experimental setup with two or three atoms and an initial coherent state
containing from 10 to 15 photons on the average. A convincing test of the generation of 
three- and four-component Schödinger cat states involving 10 photons and two or three 
atoms would require to probe the phase
distribution using an homodyne method \cite{Auffeves:2003-1}. The splitting
of the initial state into $N+1$ separated phase peaks followed by the recombination of
some of them at the time of partial spontaneous revivals would provide an experimental
proof of the scenario presented in the present paper.

A more detailed insight into the field dynamics could be gained by reconstructing the 
cavity field Wigner function, using the method proposed by and Davidovich 
and already implemented on a one-photon field in \cite{Bertet:2002-1}.

\subsubsection{Experimental consequences for circuit QED experiments}

Our analysis can be used to discuss the case of circuit-QED 
experiments \cite{Blais:2004-1}. Of course,
our model does not take into account 
relaxation nor dephasing of the atoms themselves, since they are not relevant
for the Rydberg atom experiments. In the case of circuit QED, the
relaxation and decoherence of Josephson 
qubits must be taken into account to obtain a precise model of the 
dissipative dynamics. In the litterature, the $g/\gamma $ ratio is around 20 
\cite{Wallraff:2004-1,Wallraff:2005-1}. Therefore, our model suggests that
observing the effects discussed in the present paper in circuit QED experiments 
requires an improve of roughly two orders of magnitude on this ratio.

\medskip

Increasing $g/\gamma$ would then require either an an increase of the resonator quality factor or 
an increase of the qubit/cavity coupling. Using classical eletrodynamics,
an upper bound on the value of $g$ can be estimated 
to be of the order of $g/\omega_{0}\simeq \Lambda\,
\sqrt{2\alpha/\epsilon_{r}}$ where $\alpha$ is the fine structure 
constant, $\epsilon_r$ the relative permitivity of the substrate and $\Lambda$
collects geometrical form factors and capacitance ratios determining the
microstrip impedance and its coupling to the qubit. 
For resonators in the 1-10~GHz frequency range, values of
$g/2\pi$ of the order of 100~MHz are quite realistic and would already provide 
a factor 5 on $g/\gamma$.
In present circuit-QED experiments, 
detection is performed by probing the cavity through transmission measurements.
This requires to lower the quality factor from $10^6$ \cite{Frunzio:2005-1} to $10^4$ which 
is still in the strong coupling regime but not as deeply as for Rydberg atom experiments. 
Less dissipative cavities could then be used provided one could
improve the qubits properties or use an alternative measurement method.
Recently, new detection schemes based on dynamical bifurcation of Josephson junctions
have been developed and provide high contrast, rapid measurement,
low back action and absence of on-chip 
dissipation \cite{Siddiqi:2005-1,Siddiqi:2005-2,Lupascu:2006-1}.
Thus, although an important progress is required for circuit-QED devices, the rapid improvement
of Josephson device technology is very encouraging and provides a strong
motivation for further theoretical studies. 

\section{Conclusion}

We have studied the resonant interaction of an ensemble of $N$ atoms symmetricaly coupled 
to a resonant mesoscopic field in a cavity. The interaction between the atomic ensemble and
the cavity produces an entangled state with $N+1$ components leading
to a rich pattern for Rabi oscillation revivals that generalizes the ones 
obtained in the $N=1$ case. In particular, ``fractional" spontaneous revivals
reflecting partial disentanglement of the atom + cavity state are expected to occur 
earlier than the first spontaneous revival in the $N=1$ case.

Dissipation in the cavity will lead to decoherence of
this mesoscopic entangled state and we have proposed a simple analytical
model that enables us to compute the Rabi oscillation signals in the presence of
dissipation. This model provides simple expressions for the 
spontaneous Rabi oscillations revivals as well as for the ones induced in an 
echo experiment. Analytical results are in good agreement with
quantum Monte-Carlo simulations and provide an intuitive view
of the evolution of the dissipative atoms+cavity system. We have
obtained an analytical
expression for the accumulated decoherence of the $N+1$-component Schrödinger cat 
state resulting from the atom + field interaction
which could be used for tests of decoherence. 

\medskip

We have shown that in a forthcoming generation of cQED experiments, spontaneous revivals
of Rabi oscillations associated with the recombination of a fraction of the $N+1$ components of the
entangled atoms + cavity state should be observable. 
An improvement by a factor 10 in the cavity quality factor
as well as the use of slow atoms (100 m/s) are required. Our analysis also
suggests that the situation is not so favorable within the context of circuit-QED
experiments, due to the measurement limited quality factor. However, rapid
progresses in this field are extremely encouraging and motivate further
theoretical studies taking into account relaxation and dephasing of the qubits. 
The stochastic wave function approach
could be used to take these dissipative phenomenon into account at least for the part due to 
high frequency noise. Dissipative dynamics in the presence of a strong low frequency $1/f$ noise cannot 
be accounted for within the framework of the Bloch-Redfield equations \cite{Makhlin:2004-1}.
Nevertheless, providing a simple analytical model for the dissipative dynamics of a combined
qubits + cavity system taking into account all possible sources of dissipation
would is an interesting challenge. 

\begin{acknowledgements}
We thank P.~Bertet and O.~Buisson for useful discussions on forthcoming progresses in
Josephson device technology.
\end{acknowledgements}

\appendix
\section{Mesoscopic approximation}
\label{sec:appendixA}

In the classical limit, we expect eigenstates of $J^x$ to remain unentangled with the cavity state.
Therefore, let us start from an initial state:
\begin{equation}
\label{eq:psitildeXinit}
|\Psi^X_{m}\rangle = \sum_{m''=-J}^JR^{-1}_{m'',m}
|J,m''\rangle\otimes |\alpha\rangle\,,
\end{equation}
where $\alpha = \sqrt{\bar{n}}$ and
$R_{m,m'}=\langle J,m'|e^{i\pi J^y}|J,m\rangle$. 
Since we are in the mesoscopic regime, most of the weight of
the state is concentrated in stable subspaces $\mathcal{H}_{n}$ with
$n$ close to $\bar{n}$. Apart from the term $m'=J$, the above
sum also spreads on the lower dimensional stable subspaces. Since
we expect this contribution to be exponentially weak in the mesoscopic domain, we 
focus the projection $|\tilde\Psi^{X}_{m}\rangle$
of $|\Psi^X_{m}\rangle $ in $\bigoplus_{n}\mathcal{H}_{n}$.
Because the effective Hamiltonian
\eqref{eq:Heffective} is written in terms of the $\mathcal{J}^x$ generator, 
it is useful to decompose 
$|\Psi^X_{m}\rangle $ on the basis vectors $|X_{m}^{(n)}\rangle$. We start from
\begin{equation}
|\tilde\Psi^{X}_{m}\rangle = e^{-\frac{\bar{n}}{2}}\sum_{k,m'}
\frac{\bar{n}^{k/2}}{\sqrt{k!}}\,
R^{-1}_{m,m'}
|Z_{m'}^{(k-J+m')}\rangle
\end{equation}
where the sum over $k$ ranges from $J-m'\geq 0$ to $+\infty$.
Shifting the index $k$ into  $p=k+m'-J\geq 0$ and approximating $\sqrt{\bar{n}^k/k!}$ by 
$\sqrt{\bar{n}^{p}/p!}$ enables to do the summation over $m'$
using $|Z^{(p)}_{m}\rangle=\sum_{m''}R_{m,m''}|X^{(p)}_{m''}\rangle $. 
This finally leads to:
\begin{equation}
\label{eq:psitildeXinit:adapted}
|\tilde\Psi^{X}_{m}\rangle \simeq e^{-\frac{\bar{n}}{2}}\sum_{p=0}^\infty
\frac{\alpha^{p}}{\sqrt{p!}}\,
|X_{m}^{(p)}\rangle\,.
\end{equation}
When evolved during time $t$ under the effective
Hamiltonian \eqref{eq:Heffective}, this state becomes:
\begin{equation}
|\tilde\Psi^{X}_{m}(t)\rangle = e^{-\frac{\bar{n}}{2}}
\sum_{p=0}^\infty\frac{\alpha^p}{\sqrt{p!}}\,e^{-igtm\sqrt{p+c}}\,
|X^{(p)}_{m}\rangle\,.
\end{equation}
We now use the $R^{-1}$ matrix to go back to the usual basis:
$|X_{m}^{(p)}\rangle=\sum_{m'}R^{-1}_{m,m'}\,|Z^{(p)}_{m'}\rangle$. In order to
rewrite the resulting state as a tensor product of an atomic polarization and
a field state, it is necessary to introduce the same approximations as before. We first
approximate $\sqrt{(p+l)!/p!\,\bar{n}^p}\sim 1$. The coefficient
of $|J,J-l\rangle\otimes |p+l\rangle$ is then equal to:
$$\frac{\bar{n}^\frac{p+l}{2}}{\sqrt{(p+l)!}}\,
R^{-1}_{m,J-l}e^{-igmt\sqrt{p+c}}\,.$$
We then replace $e^{-igmt\sqrt{p+c}}$ by $e^{-igmt\sqrt{p+l}}$ multiplied by
the phase factor $e^{igmt(\sqrt{p+l}-\sqrt{p+c})}$. This last phase factor
is then expanded to first order near $\bar{n}$ which leads to a slowly varying phase
$e^{igmt(l-c)/2\sqrt{\bar{n}}}$. The resulting coefficient now only depends on $l$ and
$p+l$. We then define $k=p+l$ and extend its summation range from $k=0$ to 
$+\infty$, introducing an exponentially small error in the mesoscopic regime 
$\bar{n}\gg 2J$. The resulting vector is now our mesoscopic approximation to
$|\Psi^X_{m}(t)\rangle$. It is obtained as a tensor product:
\begin{equation}
\label{eq:psitilde:factorized}
|\Psi^{X}_{m}(t)\rangle \simeq e^{-igmt\sqrt{\bar{n}}}\,|D_m(t)\rangle \otimes
|\psi_{m}(t)\rangle
\end{equation}
where $|D_{m}(t)\rangle$ and $|\psi_{m}(t)\rangle$ are defined respectively 
by eq. \eqref{eq:atomic-polarization} and \eqref{eq:GBstate}. Equation \eqref{eq:entangled-state}
is recovered by expanding $|J,m_{0}\rangle$ over the atomic polarizations at
initial time $|D_{m}(0)\rangle$ ($-J\leq m\leq J$). 

\section{Rotation matrices}
\label{sec:appendixSU2}

Matrix elements of $SU(2)$ general elements are given for example
in \cite{Arecchi:1972-1}. The 
matrix elements $R_{m,m'}=\langle J,m'|\exp{(i\pi J^y/2)}|
J,m\rangle$ needed in the present paper are:
\begin{eqnarray}
R_{m',m} & = & \sqrt{\frac{(J-m)!(J-m')!}{(J+m)!(J+m')!}}\nonumber \\
& \times &
\sum_{k=0}^{J-\mathrm{max}(m,m')}
\frac{(-1)^{J-k}(2J-k)!}{2^{J-k}k!(J-m-k)!(J-m'-k)!}\nonumber
\end{eqnarray}
Starting from excited atoms and looking for the probability of detecting finally all the atoms in the excited state involves:
$$R_{m,J}=(-1)^{J-m}R_{J,m}=\frac{1}{2^J}\,\sqrt{\frac{(2J)!}{(J-m)!(J+m)!}}$$

\section{Decoherence and phase diffusion}
\label{sec:appendix:diffusion}

In the $\gamma t\gtrsim 1$ case, the 
decay of the average photon number $\bar{n}(t)=\bar{n}\,e^{-\gamma t}$ prevents the 
sequence of quantum jumps from being a renewal process with a stationary 
waiting times probability distribution. Nevertheless, the probability that no quantum jump
happens between $t$ and $t+\tau$ knowing that one occured at time $t$ is given by:
\begin{eqnarray}
\Pi_{0}(t,\tau) & = & \prod_{t\leq t'\leq t+\tau}(1-\gamma \bar{n}(t')dt')
\nonumber\\
& = & \exp{\left(-\gamma \int_{0}^\tau\bar{n}(t+\tau')\,d\tau'\right)}\,.
\end{eqnarray}
The probability distribution for having 
quantum jumps at times $t$ and $t+\tau$ is therefore
given by 
\begin{equation}
\psi(t,\tau)=-\frac{\partial \Pi_{0}}{\partial\tau}=
\gamma\, \bar{n}(t+\tau)\,e^{-\gamma \int_{0}^\tau\bar{n}(t+\tau')\,d\tau'}\,.
\end{equation}
The probability for having exactly $p$ quantum jumps between $0$ and $t$ at times
$0\leq t_{1}\leq \ldots \leq t_{p}\leq t$ is equal to
$\mathcal{P}_{[0,t]}(t_1,\ldots ,t_p)=\psi(0,t_{1})\,\psi(t_{1},t_{2})\ldots \psi(t_{p-1},t_{p})\,
\Pi_{0}(t_{p},t-t_{p})$:
\begin{equation}
\label{eq:proba-jump-sequence}
\mathcal{P}_{[0,t]}(t_1,\ldots ,t_{p})=
e^{-\gamma \int_{0}^t\bar{n}(t')\,dt'}
\prod_{k=1}^p(\gamma \bar{n}(t_{k}))\,.
\end{equation}
Using the exponential relaxation of the
mean photon number ($\bar{n}(t)=\bar{n}e^{-\gamma t}$), this leads
to a Poisson distribution law for the number $N[0,t]$ of photons emitted between $0$ and $t$ with average
value $\bar{n}(1-e^{-\gamma t})$. Therefore, the decoherence coefficient is equal to:
\begin{equation}
\langle e^{i\theta N[0,t]}\rangle = \exp{\left(
\bar{n}\,(e^{i\theta}-1)(1-e^{-\gamma t})\right)}
\end{equation}
which is exactly the solution of the master eq. 
\eqref{eq:master-equation}. 

\section{Effect of dissipation on the atoms + cavity}
\label{sec:dissipation}

We consider the dissipative dynamics of the strongly coupled atoms + cavity systems
in the mesoscopic regime. For weak dissipation 
$\gamma \ll g$, 
we show that a family of generalized Gea-Banacloche states is stable under
an effective 
stochastic dynamics naturally arising from these approximations. We then consider the evolution
of the states $|\Psi_{m}^X\rangle$ ($-J\leq m\leq J$) and show that each of them decoheres on a time scale
much longer than the decay time of the $|\Psi_{m_+}^X\rangle\langle \Psi_{m_-}^X|$
coherence for $m_{+}\neq m_{-}$. These results validate the simplified analysis presented in
section \ref{sec:dissipative:cQED}.

\subsection{Effective stochastic dynamics in the mesoscopic regime}

%%%%%%%%%%%%%%%%%%%%%%%%%%%%%%%%%%%%%%%%%%%%%%%%%%%%%
%%%%%%%%%%%%%%%%%%%%%%%%%%%%%%%%%%%%%%%%%%%%%%%%%%%%
% j'ai le sentiment qu'il manque un i dans la partie non hermitique du H
The non hermitian Hamiltonian that describes the dynamics of the atoms + cavity system
between quantum jumps is $H_{TC}-i\hbar\gamma a^\dagger a /2$ where $H_{TC}$ is
the Tavis-Cummings Hamiltonian \eqref{eq:Dicke-SU2}. Replacing $H_{TC}$ by
the effective Hamiltonian \eqref{eq:Heffective} 
leads to an effective non hermitian Hamiltonian over each subspace
$\mathcal{H}_{p}$:
\begin{equation}
\label{eq:Heffective:dissipative}
H=\hbar g\sqrt{p+c}\,\mathcal{J}^x-i\frac{\hbar\gamma}{2}
(p+J-\mathcal{J}^z)\,.
\end{equation}
The constant term $i\hbar J\gamma /2$
can be discarded in the evolution between quantum jumps since it is canceled by 
the normalization of the state vector. 
It is then
useful to decompose the atoms + cavity vector over $\oplus_{p}\mathcal{H}_{p}$ as:
\begin{equation}
\label{eq:decomposition}
|\Psi(t)\rangle= \mathcal{N}(t)\sum_{p=0}^\infty
\frac{\alpha(t)^p}{\sqrt{p!}}\,|\Psi_{p}(t)\rangle
\end{equation}
where $|\Psi_{p}(t)\rangle$ belongs to 
$\mathcal{H}_{p}$, $\alpha(t)$ denotes some time dependent function and
$\mathcal{N}(t)$ is the normalization factor. Taking
$\alpha(t)=\alpha(0)\,e^{-\gamma t/2}$ enables to absorb the $p$-dependent part
in the non-Hermitian term of \eqref{eq:Heffective:dissipative}. As a consequence, 
each vector $|\Psi_{p}(t)\rangle$
evolves in $\mathcal{H}_{p}$ under the non-hermitian Hamiltonian:
\begin{equation}
\label{eq:Hp:nonhermitian}
H_{p}=\hbar g\sqrt{p+c}\,\mathcal{J}^x+i(\hbar \gamma/2) \mathcal{J}^z\,.
\end{equation}
The time dependence of $\alpha$ reflects the acquisition of information arising
from the absence of quantum jumps on the cavity. Because of the strong coupling
between the atoms and the cavity, the state of the atoms is also altered and this is why
dissipation has to be taken into account in the evolution of each $|\Psi_{p}(t)\rangle$
through the non-hermitian term in \eqref{eq:Hp:nonhermitian}. Note that
this induces a change in the probabilities for the cavity to release a photon
into its environment and thus modifies the cavity relaxation. 

It is possible to solve analytically the resulting Schrödinger equation
for each $|\Psi_{p}(t)\rangle$ but the resulting expressions
are very complicated. In particular, taking into account the back action of 
cavity dissipation through the atoms on the statistics of photon emission
leads to very cumbersome expressions. Nevertheless,
in the strong coupling regime of cQED ($\gamma\ll g$), neglecting
the effect of cavity dissipation on the atoms seems to be reasonable and
turns out to make the problem much more tractable. 
For values of $t$ such that $\bar{n}(t)=\bar{n}e^{-\gamma t}\gg 1$, we expect $H_{p}$ to
be dominated by its hermitian part. Discarding the
non hermitian part in \eqref{eq:Hp:nonhermitian} means
that each $|\Psi_{p}(t)\rangle$ has the same unitary evolution than in the dissipationless
case. Therefore, within this approximation, the atoms + cavity state evolves 
in the absence of quantum jumps between $0$ and $t$ as:
\begin{equation}
\label{eq:nojump-evolution}
|\Psi(t)\rangle_{\mathrm{n.j.}} = e^{-\bar{n}(t)/2}\,\sum_{p=0}^\infty
\frac{\alpha(t)^p}{\sqrt{p!}}
\,e^{-igt\sqrt{p+c}\,\mathcal{J}^x}\,|\Psi_{p}(0)\rangle
\end{equation}
where $\bar{n}(t)=|\alpha(t)|^2=|\alpha|^2e^{-\gamma t}$. 

\medskip

Let us now consider a state of the form \eqref{eq:psitildeXinit:adapted} 
and compute its evolution during a time $t$ in the absence of quantum
jump. Equation \eqref{eq:nojump-evolution} leads to: 
\begin{equation}
\label{eq:psiX:nojump:1}
|\Psi^X_{m}(t)\rangle_{\mathrm{n.j.}} =  
e^{-\bar{n}(t)/2}\,\sum_{p=0}^\infty
\frac{\alpha(t)^p}{\sqrt{p!}}
\,e^{-igmt\sqrt{p+c}}\,|X^{(p)}_{m}\rangle\,.
\end{equation}
Note that, within the mesoscopic approximation, this state can be approximated by
a factorized state of the form \eqref{eq:psitilde:factorized} taking into account dissipation through
the exponential decay of $\bar{n}(t)$. For $\gamma t\ll 1$, eq. \eqref{eq:psitilde:factorized} 
provides a good approximation to \eqref{eq:psiX:nojump:1}.

\medskip

We now consider the effect of quantum jumps on state \eqref{eq:psiX:nojump:1}.
Within the mesoscopic approximation, the action of the creation operator on states
$|X_{m}^{(p)}\rangle$ can be simplified:
\begin{eqnarray}
a\,|X_{m}^{(p)}\rangle & = & 
\sqrt{p+J-m}\,|J,m\rangle\otimes |p+J-m-1\rangle \nonumber \\
& \simeq & \sqrt{p}\,|X_{m}^{(p-1)}\rangle\,.
\end{eqnarray}
Using this expression, we see that $a\,|\Psi^X_{m}(t)\rangle_{\mathrm{n.j.}}$
has the same form than \eqref{eq:psiX:nojump:1}, the 
phase $e^{-igmt\sqrt{p+c}}$ in front of $|X^{(p)}_{m}\rangle$
being replaced by $e^{-igmt\sqrt{p+1+c}}$. This shows that all states of the
form:
\begin{equation}
\label{eq:dissipation:stable-states}
|\Psi_{m}[\alpha,\lbrace\Phi(t)\rbrace]\rangle=e^{-\frac{\bar{n}(t)}{2}}
\sum_{p=0}^\infty
\frac{\alpha (t)^p}{\sqrt{p!}}
\,e^{-i\phi(p,t)}\,|X^{(p)}_{m}\rangle\ ,
\end{equation}
where $\Phi(t)$ denotes the set of all phases $\phi(p,t)$ ($p\geq 0$) at time $t$,
form a stable class along the stochastic evolution. The phases $\phi(p,t)$ follow a stochastic
trajectory consisting of smooth deterministic evolution periods between quantum jumps. 
The deterministic evolution is ruled
by $\dot{\phi}(p,t)=mg\sqrt{p+c}$. The quantum jumps correspond to discontinuous 
steps $\phi(p,t^+)=\phi(p+1,t)$. 

\subsection{Evolution of the states $|\Psi_{m}^X\rangle$}

Studying the evolution of a superposition of states of the form \eqref{eq:psitildeXinit:adapted} 
leads to consider the evolution of a coherence between $|\Psi_{m_{+}}^X\rangle$ and
$|\Psi_{m_{-}}^X\rangle$. After time $t$, the stochastic evolution of
$|\Psi_{m_+}^X\rangle\langle \Psi_{m_-}^X|$ produces an ensemble
of projectors of the form: $|\Psi_{m_+}[\alpha,\lbrace\Phi_{+}(t)\rbrace]\rangle
\langle\Psi_{m_-}[\alpha,\lbrace\Phi_{-}(t)\rbrace]|$. Denoting
by $\rho_{m_+,m_-}(t)$ the operator obtained by averaging these projectors
over the measure given by the stochastic trajectories, we obtain:
\begin{eqnarray}
\rho_{m_+,m_-}(t) & = & e^{-\bar{n}(t)}\sum_{(p_{-},p_{+})}
\frac{\alpha(t)^{p_-+p_+}}{\sqrt{p_-!\,p_+!}}\nonumber\\
& \times &
D_{p_{-},p_{+}}(t)\,|X_{m_{+}}^{(p_{+})}\rangle
\langle X_{m_{-}}^{(p_{-})}|\,.
\end{eqnarray}
where $D_{p_{+},p_{-}}(t)$ denotes the average over all stochastic trajectories of the
relative phase factor $e^{i(\phi_{+}(p_{-},t)-\phi_{-}(p_{+},t))}$ where
the phases $\phi_{\pm}(p,t)$ are relative to the states $\Psi_{m_{\pm}}$. 
Because the  quantum jump process is memoryless, these averages obey the
following set of coupled first order differential equations:
\begin{eqnarray}
\label{eq:Dppluspminus:dynamics}
\dot{D}_{p_+,p_-}(t) & = &  \gamma\,\bar{n}(t)\,
(D_{p_++1,p_-+1}(t)-D_{p_+,p_-}(t))\nonumber \\
& - & i\Omega_{p_-,p_+}^{(m_{-},m_{+})}\,D_{p_+,p_-}(t)\,.
\end{eqnarray}
where
\begin{equation}
\Omega_{p_+,p_-}^{(m_{-},m_{+})}=g\,(m_{+}\sqrt{p_++c}-m_{-}\sqrt{p_-+c})\,.
\end{equation}
Let us first focus on the case $m_{+}=m_{-}=m$. 
Probing the
$(p_{+},p_{-})$ dependence of $D_{p_{+},p_{-}}(t)$ gives us an insight of the decoherence
of the initial pure state $|\Psi_{m}^X\rangle$ of the atoms + cavity system. 
Decoherence arises
from the quantum jumps that lead to the spreading of $\phi(p_{+},t)-\phi(p_{-},t)$. Let us estimate
the decoherence coefficient between $|X^{(p_+)}_{m}\rangle$ and $|X^{(p_-)}_{m}\rangle$ within the
mesoscopic approximation. Expanding $\sqrt{p+c+1}\simeq\sqrt{p+c}+1/2\sqrt{p}$
the phase factor associated with a single jump at time $t_j$ is  
$gmt_j/2\sqrt{p_{+}}-gmt_{j}/2\sqrt{p_{-}}$
which is approximately equal to $gmt_j(p_{-}-p_{+})/4\bar{n}^{3/2}$ for $p_{\pm}$ close to
$\bar{n}$ (we assume $\gamma t_j\lesssim 1$ for simplicity). Thus, for $m_{+}=m_{-}=m$,
the decoherence factor $D_{p_+,p_-}(t) $ can be approximated by
\begin{equation}
D_{p_+,p_-}(t)\simeq e^{-igmt(\sqrt{p_++c}-\sqrt{p_-+c})}
\langle e^{i \frac{p_+-p_-}{2\,\bar{n}}\sum_{j}\frac{gmt_j}{2\sqrt{\bar{n}}}}
\rangle
\end{equation}
Remembering that the statistics of occurrence times of quantum jumps for all states of the
form \eqref{eq:dissipation:stable-states} is computed in appendix
\ref{sec:appendix:diffusion}, we immediately obtain ($\gamma t\lesssim 1$)
\begin{eqnarray}
D_{p_+,p_-}(t) & \simeq & e^{-igmt(\sqrt{p_++c}-\sqrt{p_-+c})}\nonumber\\
& \times &
e^{\bar{n}\gamma \int_{0}^t(e^{i\eta_{p_{+},p_{-}}\theta_{m}(\tau)}-1)\,d\tau}
\label{eq:Dmm}
\end{eqnarray}
where $\theta_{m}(\tau)=gm\tau/2\sqrt{\bar{n}}$ and $\eta=(p_{+}-p_{-})/2\bar{n}$.
The second factor in \eqref{eq:Dmm} is responsible for decoherence of the state $|\Psi^X_{m}\rangle$
of the atoms + cavity system. Because of the amplitude $e^{-\bar{n}(t)/2}
\frac{\alpha(t)^{p_{\pm}}}{\sqrt{p_{\pm}!}}$
in the atoms + cavity states, the values of $p_{\pm}$ that contribute to the sum
lie within $|p_{+}-p_{-}|\lesssim \sqrt{\bar{n}(t)}$ and therefore $|\eta|\ll 1$
within the mesoscopic regime and for $\gamma t\lesssim 1$. That's why decoherence of a state
\ref{eq:psitildeXinit:adapted} can be neglected as in \cite{Banacloche:1993-1}. 

Before moving on the $m_{+}\neq m_{-}$ case, it is interesting to see how his results (section III.A) 
are recovered within the present approach. Following \cite{Banacloche:1993-1}, we ignore the discrete
character of $p$ and replace the finite difference equation \eqref{eq:Dppluspminus:dynamics} 
by partial differential equation.  Solving this equation can easily be done using the characteristics 
method. Starting from the initial condition $|\Psi_m^X\rangle$, this leads to 
$\rho_{m,m}(t)=|\Psi_{m}(t)\rangle\langle\Psi_{m}(t)|$ where:
\begin{equation}
|\Psi_{m}(t)\rangle=e^{-\frac{\bar{n}(t)}{2}}
\sum_{p=0}^\infty
\frac{\alpha (t)^p}{\sqrt{p!}}
\,e^{-i\Theta(p,t)}\,|X^{(p)}_{m}\rangle
\end{equation}
and
\begin{equation}
\Theta(p,t)=gm\int_{0}^t
\sqrt{c+p+\bar{n}(\tau)-\bar{n}(t)}\,d\tau\,.
\end{equation}
Evaluating the integral and for all $p$ leads to
expressions corresponding to eqs. (18.a) to (18.c) of \cite{Banacloche:1993-1}.

\medskip

The case $m_+\neq m_-$ can then be studied along the same lines.
Within the mesoscopic approximation, the phase jump associated with a quantum jump occuring at
time $t_j$ can be evaluated as 
$\exp{(igt_j(m_{-}-m_{+})/2\sqrt{\bar{n}(t_j)})}$. Note that it does not vanish for
$p_+=p_-\simeq \bar{n}(t_{j})$. This is why, decoherence for $m_+\neq m_-$
occurs on a much shorter time
scale than the decoherence of the state $|\Psi^X_{m}\rangle$. 
In principle, $\rho_{m_{+},m_{-}}(t)$ could be computed
from the formalism presented here but this is not necessary for the present
purpose. In the $\gamma t\ll 1$ case, the problem can be simplified
by considering that the evolution of the atoms + cavity state with initial
condition $|\Psi^X_m\rangle$ produces a pure state as in the previous paragraph and
by approximating this pure state by $|\Psi^X_{m}(t)\rangle$ (see 
eq.~\eqref{eq:psitilde:factorized:main}).
The argumentation presented in section \ref{sec:dissipative:cQED} then 
leads to the decoherence properties of the atoms + cavity state suitable
for this regime.

% Bibliography

\end{document}